\documentclass[onecolumn]{article}
\usepackage[english]{babel}

\usepackage[a4paper,top=2cm,bottom=2cm,left=3cm,right=3cm,marginparwidth=1.75cm]{geometry} 

\usepackage{authblk}
\usepackage{amsmath}
\usepackage{graphicx}
\usepackage{enumerate}
\usepackage[normalem]{ulem}
\usepackage[linkcolor = blue, citecolor = blue, urlcolor = blue, colorlinks = true]{hyperref}
\usepackage{bm}
\usepackage[separate-uncertainty]{siunitx}
\usepackage{amssymb}
\usepackage{lineno}

\title{\textbf{\Large Designing highly efficient lock-and-key interactions in anisotropic active particles}}

\author[1]{Solenn Riedel}
\author[2]{Ludwig A. Hoffmann}
\author[2]{Luca Giomi}
\author[1]{Daniela J. Kraft}

\affil[1]{Soft Matter Physics, Huygens-Kamerlingh Onnes Laboratory, Leiden University, PO Box 9504, 2300 RA Leiden, the Netherlands}
\affil[2]{Instituut-Lorentz, Leiden University, P.O. Box 9506, 2300 RA Leiden, The Netherlands}

\begin{document}

\maketitle

\begin{abstract}
Cluster formation of microscopic swimmers is key to the formation of biofilms and colonies, efficient motion and nutrient uptake, but, in the absence of other interactions, requires high swimmer concentrations to occur. Here we experimentally and numerically show that cluster formation can be dramatically enhanced by an anisotropic swimmer shape. We analyze a class of model microswimmers with a shape that can be continuously tuned from spherical to bent and straight rods. In all cases, clustering can be described by Michaelis-Menten kinetics governed by a single scaling parameter that depends on particle density and shape only. We rationalize these shape-dependent dynamics from the interplay between interlocking probability and cluster stability. The bent rod shape promotes assembly even at vanishingly low particle densities and we identify the most efficient shape to be a semicircle. Our work provides key insights into how shape can be used to rationally design out-of-equilibrium self-organization, key to creating active functional materials and designing targeted two-component drug delivery.
\end{abstract}

\begin{figure}
    \centering
    \includegraphics[width=0.5\textwidth]{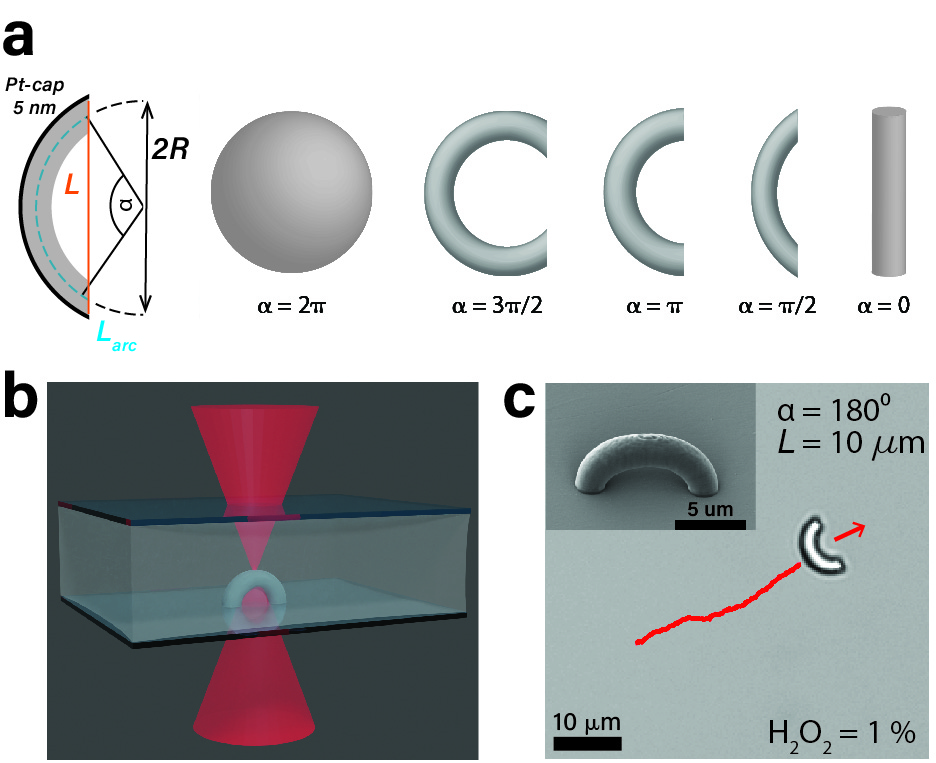}
    \caption{\textbf{Experimental design of self-propelled bent rods} (a) By changing the design parameters, the shape of the anisotropic microswimmers can be continuously tuned from a sphere to a bent or straight rod. Cross-sectional length $L$, arclength $L_{arc}$, opening angle $\alpha$, and radius of curvature $R$. (b) Schematic illustration of 3D printing of a bent rod by two-photon polymerization. (c) Scanning electron microscopy image (top left inset) and bright-field image of a 3D printed active crescent with $\alpha=\ang{180}$. The particle moves, concave-side leading, in the direction of the red arrow. Red line is trajectory of past 30 s.}
    \label{Fig_1: crescent clustering scheme}
\end{figure}

\begin{figure}
\centering
\includegraphics[width=0.95\textwidth]{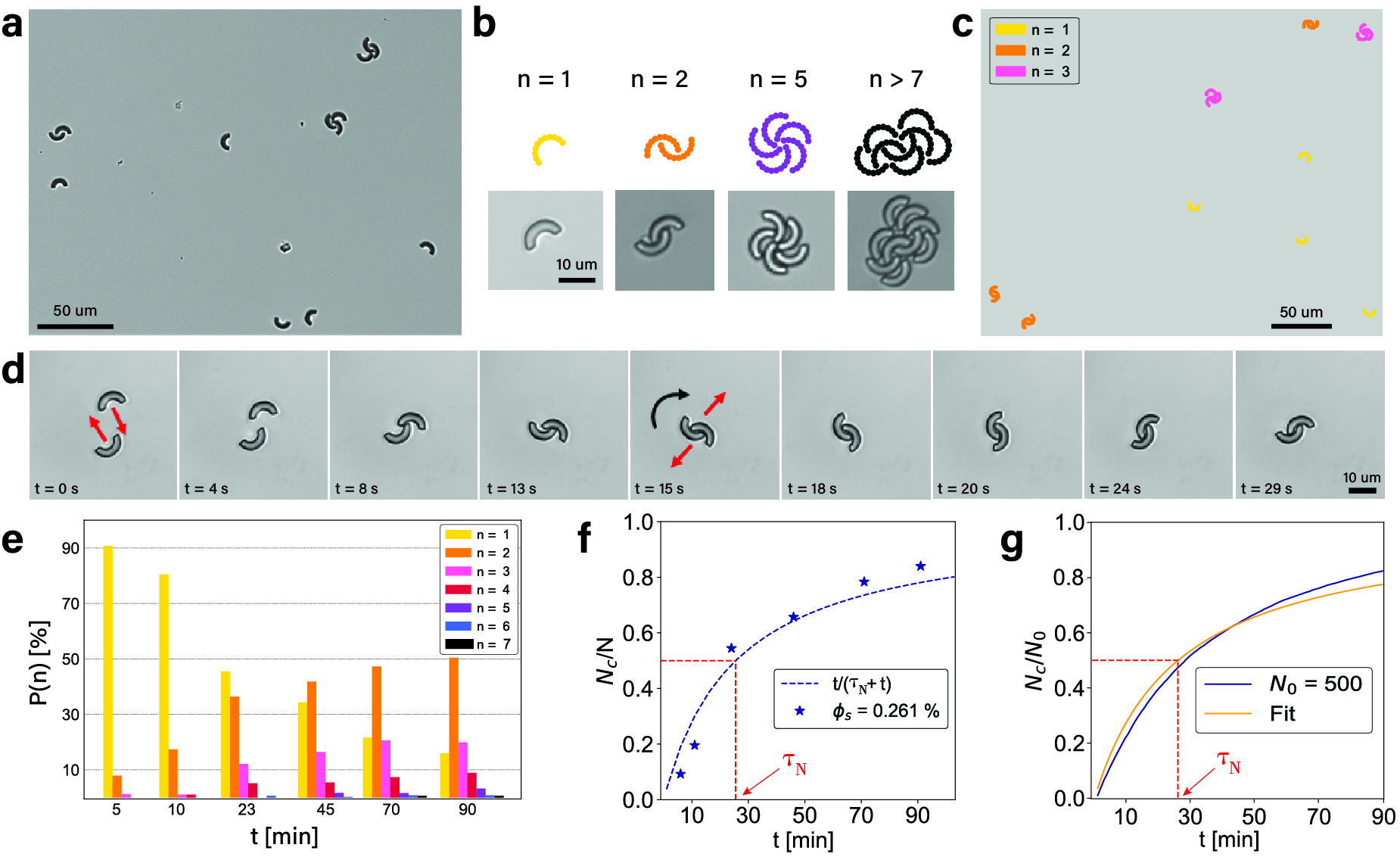}
\caption{\textbf{Time-dependent clustering of active crescents} 
(a) Bright-field microscopy image taken 90 min after mixing particles and fuel solution shows that interlocked pairs and clusters, as well as single particles are present (surface-area fraction $\phi_s=0.261\%$, corresponding to 104 crescents/\si{\milli\meter\squared}). (b) Simulation snapshots and bright-field microscopy images of clusters of different size $N$. (c) Snapshot of simulations for $\phi_s=0.261\%$. Clusters are colored according to size. (d) Image sequence of formation of a 180\textdegree{} crescent-pair in time. The pair only rotates after interlocking. (e) Time evolution of the probability $P(n)$ to find a crescent in a cluster of size $n$ ($\phi_s=0.261\%$, measured after 90 mins). (f) Time evolution of fraction of particles that are clustered, $N_c/N$, for the data shown in (e). Eq.~\eqref{eq:FitFuncFromBiB} was used as a fit with $\tau_N = 25.5\pm4.0\min$. (g) Balls-into-bins model with $N=500$ and a system size comparable to experimental setup behaves similar to experiments and simulations. Fit of Eq.~\eqref{eq:FitFuncFromBiB} with  $\tau_N = 26.2\pm0.24\min$. Details in SI.}
\label{Fig_2: time dependence}
\end{figure}

The motility inherent to synthetic~\cite{Palacci2013, Theurkauff2012, bricard_emergence_2013} and living~\cite{Peruani2012,Petroff2015,Tan2022} microswimmers is crucial for their collective behavior. Besides pushing these systems out of equilibrium, it is at the heart of the occurrence of trains~\cite{tung_fluid_2017, Ketzetzi2022}, flocks~\cite{bricard_emergence_2013}, vortices~\cite{bricard_emergent_2015}, and clusters~\cite{ Palacci2013,Theurkauff2012,Peruani2012,Petroff2015,Tan2022,Tailleur2008}. Cluster formation occurs in a range of active agents, from self-propelled colloids forming living active crystals to rotating or moving clusters of populations of starfish embryos~\cite{Tan2022}, or motile bacteria like \textit{Thiovulum majus} \cite{Petroff2015} and \textit{Myxococcus xanthus} mutants \cite{Peruani2012}. There, it is thought to be important for the formation of colonies and biofilms, efficient motion, and nutrient uptake~\cite{solari_motility_2007}. 
\\
Simulations of self-propelled agents have identified that motility is a minimal requirement for cluster formation, and that it occurs even in systems of repulsive agents at sufficiently high densities. This so-called \textit{motility induced phase separation} (MIPS) \cite{Tailleur2008, Bialke2012, Cates2015, Fily2014, Marchetti2016} originates from a slowing-down of the active agents upon collision. Systems of self-propelled disks phase separate above a critical density of 40\% \cite{Fily2012}, forming a single, globally disordered, macroscopic MIPS-aggregate that experiences diffusive motion. In experiments with synthetic microswimmers, however, cluster formation usually sets in already at several percent due to the presence of additional attractive interactions, and multiple disordered and dynamic aggregates are found \cite{Palacci2013, Theurkauff2012, Buttinoni2013}.
\\
In contrast to synthetic active particles, biological microswimmers often have non-spherical shapes \cite{Schuech2019, Muthinja2017, Spreng2019, Young2006} and simulations have predicted that their collective behavior is strongly influenced by shape. For example, longitudinally-propelled rods self-organize into polar bands instead of clusters as a consequence of the lateral association promoted by their elongated shape~\cite{Grossmann2020, Ginelli2010, Abkenar2013, VanDamme2019}. Clusters of anisotropic active particles may not only display Brownian diffusion, but also directed and spinning motion~\cite{Wensink2014, Moran2022}. In addition,
the critical density at which cluster form has been proposed to depend on shape because it strongly influences inter-particle slowing down~\cite{Moran2022}. For active hexagons, for example, efficient deceleration upon collision results in nucleation of many small clusters at far lower particle densities than would be expected for spheres \cite{Moran2022}. 
\\
Despite these predictions, there is little experimental work due to the limited availability of active particles with an anisotropic shape. Experiments with transversely-propelled rods revealed that stable doublets already form at particle densities of $\approx$1\% \cite{Vutukuri2016} and clustering is favored with increasing aspect ratio of the rods~\cite{Peruani2006, Vutukuri2016}. Tori, both horizontally and vertically oriented, were observed to form dynamic unstable clusters~\cite{Baker2019} and standing disks were found to cluster at about 10\%~\cite{Katuri2022}. While more complex shapes have been prepared~\cite{shelke_transition_2019,wang_active_2019} it is often challenging to selectively generate sufficient quantities to test their phase behavior and typically not possible to gradually tune their shape to find optimal clustering conditions.  
\\
Here, we use 3D micro-fabrication to create active particles with a shape that is continuously tunable between a sphere and a rod, i.e. bent rods, and study their collective behavior. We vary the density to below $0.022\%$ and find that clustering still occurs at such extremely low surface area fractions due to a highly efficient nucleation process. Complementing our experiments with simulations and an analytical model, we find that the self-organization can be captured by a Michaelis-Menten kinetics and thus can be characterized by a single scaling parameter that depends on the particle density and shape only. We demonstrate that the efficiency of the self-organization process is strongly influenced by the concave shape, which affects the interplay between interlocking and stability. Our insights provide a generic understanding of how phase separation occurs in a whole class of anisotropic particles from spheres to rods as well as a strategy to precisely control the stability of active particles through the shape of their interaction site, thus enabling the rational design of their assembly pathways into functional active materials.  
\\
\\
\textbf{Self-propelled bent rods}\\
We exploit 3D microprinting based on two-photon polymerization to create active particles with a tunable anisotropic shape. We print bent rods with varying opening angle $\alpha$ and constant cross-section $L = 10$~\si{\micro\meter} as this shape interpolates smoothly between a sphere and a rod, see Fig.~\ref{Fig_1: crescent clustering scheme}a, b~\cite{Doherty2020}. We render the particles active by sputter coating them on their convex side with a 5 nm thick Pt/Pd (80/20) layer and dispersing them in aqueous hydrogen peroxide (H$_2$O$_2$) solution. Their self-propulsion is driven by solute gradients generated through a catalytic decomposition of H$_2$O$_2$ at the  Pt/Pd cap \cite{Howse2007}. Depending on the fuel concentration, the bent rods either swim with their concave (1\% H$_2$O$_2$) or convex side (5\% H$_2$O$_2$) leading, see Supplementary Videos 1-3. Due to their size, the motion of these active crescents shows long persistence lengths, see Fig.~\ref{Fig_1: crescent clustering scheme}c. 
\\
\\
\textbf{Collective Behavior of Active Bent Rods}\\
We start by examining the collective behavior of active bent rods that swim with their concave-side leading, have an opening angle of $\alpha=180$\textdegree{} and an average velocity of $\langle v\rangle_{180} = 0.78\pm0.08$~\si{\micro\meter\per\s}. Small rotating clusters and pairs of particles quickly form after sedimentation, making our system effectively 2D, see Fig.~\ref{Fig_2: time dependence}a,b and Supplementary Videos 4-6. Simulations using a minimal model that only contains two ingredients, i.e., self-propelling bent rods and steric interactions between particles, \cite{Wensink2012}, also show the quick formation of stable pairs and rotating clusters (Fig.~\ref{Fig_2: time dependence}c).
Clearly, the concave shape of the active particles is crucial for clustering as it promotes interlocking and stability against breakup. 
\\
Remarkably, stable interlocking occurs already for only two particles upon a head-on collision. The translational motion of the individual particles is then transformed into a stationary rotation of the pair, see Fig.~\ref{Fig_2: time dependence}d. The direction of rotation is determined by the torque created by the initial off-center alignment of the particles and remains stable over the duration of our experiment, with a typical angular velocity of $\approx 0.13\pm0.01$ \si{\radian\per\s}. The quick formation and long stability of crescent-pairs seen in both experiments and simulations is very different from the behavior of pairs of active spheres, which, because of a lack of aligning interaction, tend to slide pass each other as soon as their velocities depart from a perfectly anti-aligned configuration. While cluster formation of spheres requires at least three, but typically more, particles~\cite{Buttinoni2013,Palacci2013,Theurkauff2012}, the crescent shape favors  configurations in which the end of one crescent is locked at the center of another particle and thus already stabilizes clusters of two particles. 
\\
The stable, rotating pairs subsequently act as nucleation points for larger clusters. Notably, clusters only grow through addition of single crescents and not by cluster merging, because of a lack of translational motion once the crescents form a pair. The long lifetime of the pairs enhances nucleation, similar to active polygons studied in simulations~\cite{Moran2022}. But unlike active polygons, active crescents can interlock as their shape features a cavity. This makes their self-assembly stable against break-up, even upon a collision with a free particle, which is often absorbed into the cluster owing to its concave shape. We quantify the evolution of the cluster distribution over time and show the result for concave-side leading active bent rods (at $\alpha =\ang{180}$, 1\% H$_2$O$_2$) in Fig.~\ref{Fig_2: time dependence}e. Already after 10~min about 17\% of crescents have formed pairs, despite our experiments taking place at a very low particle density of only $\phi_s=0.261\%$, which corresponds to only about 104 particles/mm$^2$. Over the next 80~min the amount of single particles drops to 16\%, more than half the particles are paired, while three- and four-particle clusters form in smaller amounts (20\% and 9\%, respectively).
\\
This efficient nucleation and growth process leads to a quick assembly of the active bent rods into clusters. We find that the time evolution of the fraction of particles that are part of a cluster, $N_c$, relative to the total number of particles in the system, $N$, quickly grows such that the majority of particles are part of a cluster after only about 30 min, i.e. $N_c/N >0.5$ (Fig.~\ref{Fig_2: time dependence}f), and almost all particles are part of a cluster after 90~min (Fig.~\ref{Fig_2: time dependence}e) despite the very low particle density of $\phi_s=0.261\%$.
\begin{figure*}
\centering
\includegraphics[width=0.95\textwidth]{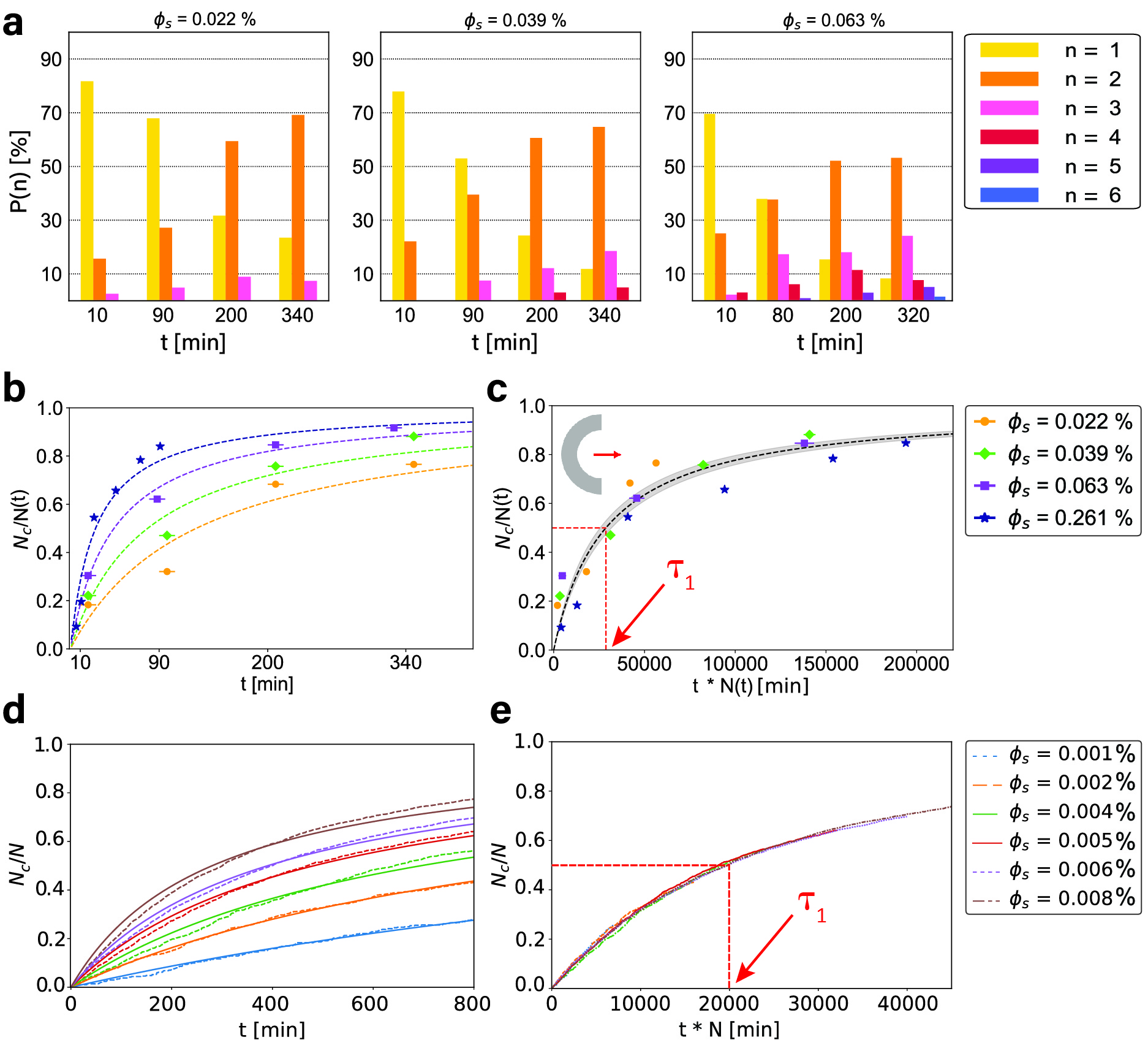}
\caption{\textbf{Density dependent clustering} of concave-side leading, bent rods ($\alpha=\ang{180}$, $L = 10$~\si{\micro\meter}). (a) Cluster-size distribution over time for different surface-area fractions $\phi_s$, corresponding to particle number densities of (from left to right) 9, 15, and 25 crescents/\si{\milli\meter\squared}. Note that $\phi_s$ is time-dependent and the stated values are measured after 90 min (see main text and SI). 
(b) $N_c/N(t)$ for different densities fitted with Eq.~\eqref{eq:FitFuncFromBiB}. The observation time needed to cover the entire sample reflects as an $x$-error for each data point (also applicable to (c)). (c) Master curve obtained from (b) by rescaling time as $t\to t N(t)$. Fit with Eq.~\eqref{eq:FitFuncFromBiB} with  $\tau_1 = (2.9\pm0.3)10^4\min$. (d) Cluster-size distribution over time from simulations (average of 100 runs). (e) $N_c/N(t)$ from simulations for various densities, and (f) the corresponding master curve, where $\tau_1 = (1.99\pm0.10)10^4\min$.}
\label{Fig_3: concentration dependence}
\end{figure*}
To better understand the observed clustering process we consider a balls-into-bins model, where particle clustering is modelled as the presence of multiple particles (balls) inside a finite set of containers (bins) after random assignment (see Methods and SI). Each time step, all single particles are randomly assigned a new bin. A $N$-particle cluster exists if a bin contains $N$ particles. Simulating many time steps, we find a $N_c/N$ curve as shown in Fig.~\ref{Fig_2: time dependence}g, which is qualitatively similar to those of the experiments.
\\
To obtain an analytical expression, we consider only two-particle-cluster formation for simplicity, because dimers are (initially) dominant. We then find that the time evolution of the average number of dimerized particles can be described by a Poisson process $\langle N_c (t) \rangle = 2 r t$ with a \emph{time-dependent} rate $r=r(t)$. Assuming a linear decrease of non-dimerized particles, $N_f=N - 2 r t$, we find the Michaelis-Menten equation
\begin{equation}
\frac{\langle N_{c}\rangle}{N} = \frac{t}{\tau_{N} + t} \;,
\label{eq:FitFuncFromBiB}
\end{equation}
where $\tau_{N}$ is the time required for two particles to collide and interlock (see Methods and SI). At short times $\langle N_c \rangle/N \sim t/\tau_{N}$ increases linearly with $t$, as the abundance of non-dimerized crescents makes the rate approximately time-independent. At long times, on the other hand, all particles are dimerized and $\langle N_c \rangle/N \to 1$. Fitting Eq.~\eqref{eq:FitFuncFromBiB} to our experimental and numerical data we find very good agreement, as confirmed by Fig.~\ref{Fig_2: time dependence}f, g and Fig. \ref{Fig_3: concentration dependence}b. In the following we will return to using the model of Ref.~\cite{Wensink2012} for simulations.
\\
\\
\textbf{Density Dependence of Crescent Clustering}\\ 
Our initial results demonstrated that significant clustering already appears at very low particle densities. To investigate the density dependence of clustering and explore whether there is a limit, we tested three even lower particle concentration in experiments. Remarkably, even at concentrations as low as nine crescents/$\text{mm}^{2}$ (i.e. $\phi_s=0.022\%$) cluster formation still happens, see Fig. \ref{Fig_3: concentration dependence}a, albeit slower. This surface area fraction is two orders of magnitude smaller than for active spheres~\cite{Palacci2013,Theurkauff2012,Fily2012}, rods~\cite{Vutukuri2016}, and standing disks~\cite{Katuri2022}. 
We also return to simulations to investigate even lower particle densities. Similar to the experiments, a significant number of particles cluster (Fig. S11). Plotting $N_c/N$ in Fig.~\ref{Fig_3: concentration dependence}b,d, we find that the assembly occurs slower at lower particle densities, but still looks qualitatively similar. The Michaelis-Menten Eq. \eqref{eq:FitFuncFromBiB} fits all the simulation and experimental data well, independent of the density. 
\\
Eq.~\eqref{eq:FitFuncFromBiB} suggests the presence of only a single time scale, the constant $\tau_{N}$, for cluster formation. This is surprising since the active Brownian motion of isolated particles, as well as the interaction between particles could, in principle, give rise to other time scales. To explain this we assume that particle collision occurs predominantly in the diffusive regime, where the particles' mean squared displacement is given by $\langle\Delta\bm{r}(t)\rangle^{2}=4\mathcal{D}t$, with $\mathcal{D}$ the diffusion coefficient. Thus, the time for two particles to collide is approximately $\tau_{N} \sim \ell^{2}/(4\mathcal{D})$, with $\ell$ the average inter-particle distance. If particles are uniformly distributed in space, $(D/\ell)^{2} \approx N$, with $D$ the system size, thus $\tau_{N} \approx D^{2}/(4\mathcal{D}N)$. Notice that, consistently with our combinatoric calculation, $\tau_{N}\sim 1/N$  (see Methods). Transforming $t \to tN$ is then equivalent to rescaling time by the only time scale of the process, thus removing any particle-density dependence.
\\
Indeed, with this simple rescaling of time, we obtain a master curve for the data from experiments and simulations described by a single fit parameter $\tau_1$, shown in Fig. \ref{Fig_3: concentration dependence}c, e. We attribute the difference in fit parameter $\tau_{1}=N \tau_N$ of approximately an order of magnitude to flow-induced attractions present in experiments as well as to slow sedimentation of particles which increases the surface concentration in time, see SI for a discussion. Both effects only modify the time scale but are not relevant for the basic clustering dynamics. We can thus conclude that the dimerization of active crescents is a single-time-scale process, regardless of the particle density. Furthermore, clustering occurs even at arbitrarily low concentrations. This is in stark contrast to spheres, which undergo a dynamic clustering process that consists of break up and growth, and the transient pairs formed by rods and disks. This peculiarity of active crescents ultimately determines the high performance of their assembly and originates from the efficiency of the lock-and-key mechanism governing dimerization. This implies an infinite interaction time and, effectively, a vanishing ballistic-diffusive crossover time thus leaving $\tau_{N}$ as the only finite time scale of the process.  
\\
\\ 
\textbf{Shape Dependence of Crescent Clustering}\\
\begin{figure*}
\centering
\includegraphics[width=0.7\textwidth]{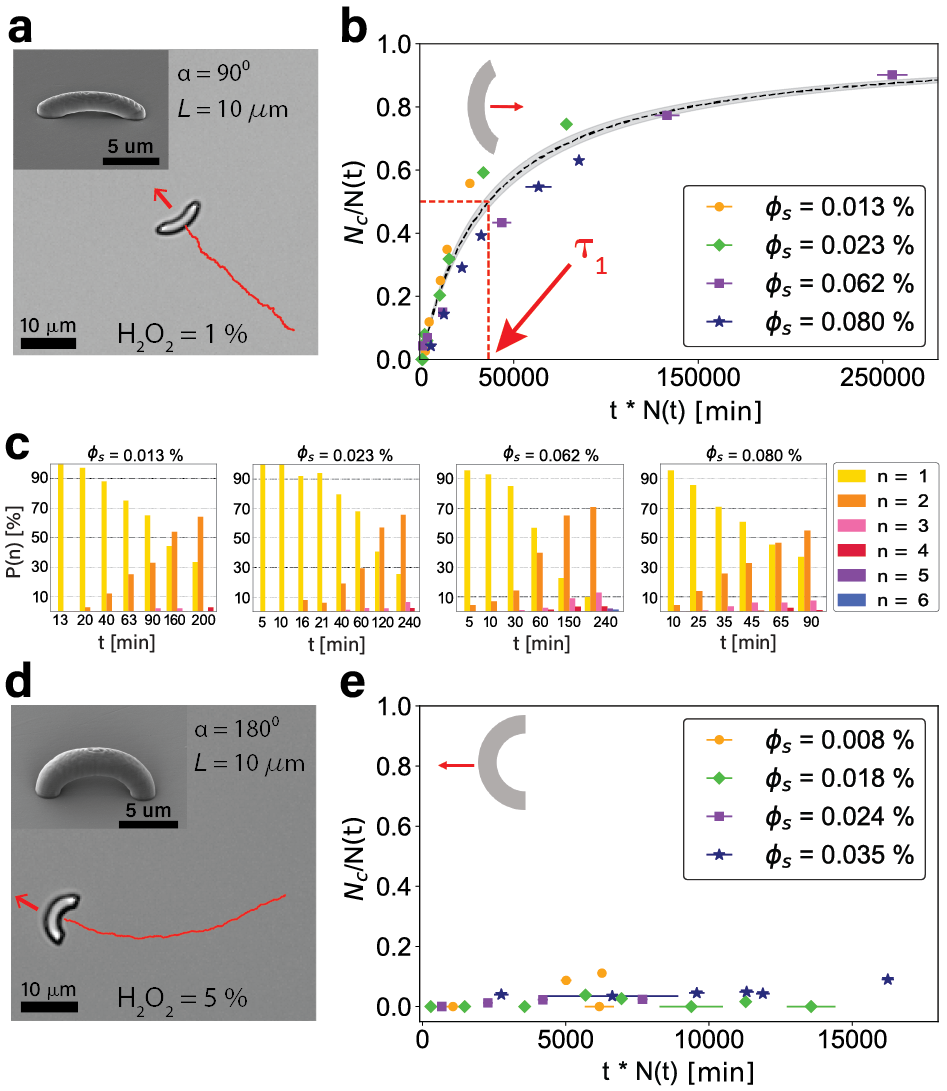}
\caption{\textbf{Shape-dependent clustering behavior} (a) Scanning electron microscopy and bright-field image of a 3D printed crescent with $\alpha=\ang{90}$. The trajectory travelled by the active crescent within 30 s is shown in red. The red arrow indicates the direction of motion, the crescent moves concave-side leading. (b) Master curve for samples with $\alpha=\ang{90}$ crescents with different surface area fractions $\phi_s$ which correspond to particle densities (after 60 min) of 7, 12, 32, and 42 crescents/\si{\milli\meter\squared}, respectively. The fit parameter obtained from fitting with Eq.~\eqref{eq:FitFuncFromBiB} is $\tau_1 = (3.63\pm0.31)10^4\min$. The observation time needed to cover the entire sample reflects an $x$-error for each data point (also applicable to (e)). (c) Cluster-size distribution over time corresponding to (b). $\ang{90}$ crescents form fewer higher-order clusters than their $\ang{180}$ counterparts at the same concentration and time. (d) Scanning electron microscopy and bright-field image of a 3D printed crescent with $\alpha=\ang{180}$. The trajectory travelled by the active crescent within 30 sec is shown in red. The red arrow indicates the direction of motion, the crescent moves convex-side leading. (e) Fraction of particles in a cluster plotted after rescaling time as $t\to t N(t)$ for systems of convex-side leading crescents ($\alpha=\ang{180}$) with different surface area fractions $\phi_s$. These correspond to particle densities (after 50 min) of 3, 7, 10, and 14 crescents/\si{\milli\meter\squared}, respectively.}
\label{Fig_4: shape dependence}
\end{figure*}
To study the connection between shape and collective behavior we print a second type of bent rod with the smaller opening angle $\alpha=\ang{90}$, keeping $L$ constant, see Fig.~\ref{Fig_4: shape dependence}a. We measured their clustering behavior for four surface area fractions ranging from $\phi_s=0.013$\% to $\phi_s=0.080$\%, and find that the $N_c/N$ data can again be rescaled into a single master curve. The fit parameter $\tau_1$ increases from $\tau_1 = (2.87\pm0.31)10^4\min$ for $\alpha=\ang{180}$ (Fig~\ref{Fig_3: concentration dependence}c) to $\tau_1 = (3.63\pm0.31)10^4\min$ for $\alpha=\ang{90}$ (Fig.~\ref{Fig_4: shape dependence}b), reflecting the slower assembly process.  
\\
This difference is even more noteworthy as the measured average speed for 90\textdegree{} crescents,  was $\langle v\rangle_{90} = 1.02\pm0.03$~\si{\micro\meter\per\s} at 1\% H$_2$O$_2$, which is higher than the speed of 180\textdegree{} particles, $\langle v\rangle_{180} = 0.78\pm0.08$~\si{\micro\meter\per\s}. Moreover, the number density of particles in the sample is higher at the same $\phi_s$ for the 90\textdegree{} crescents, due to their shorter arc-length, underlining the difference further.
\\
\begin{figure*}
\centering
\includegraphics[width=0.5\textwidth]{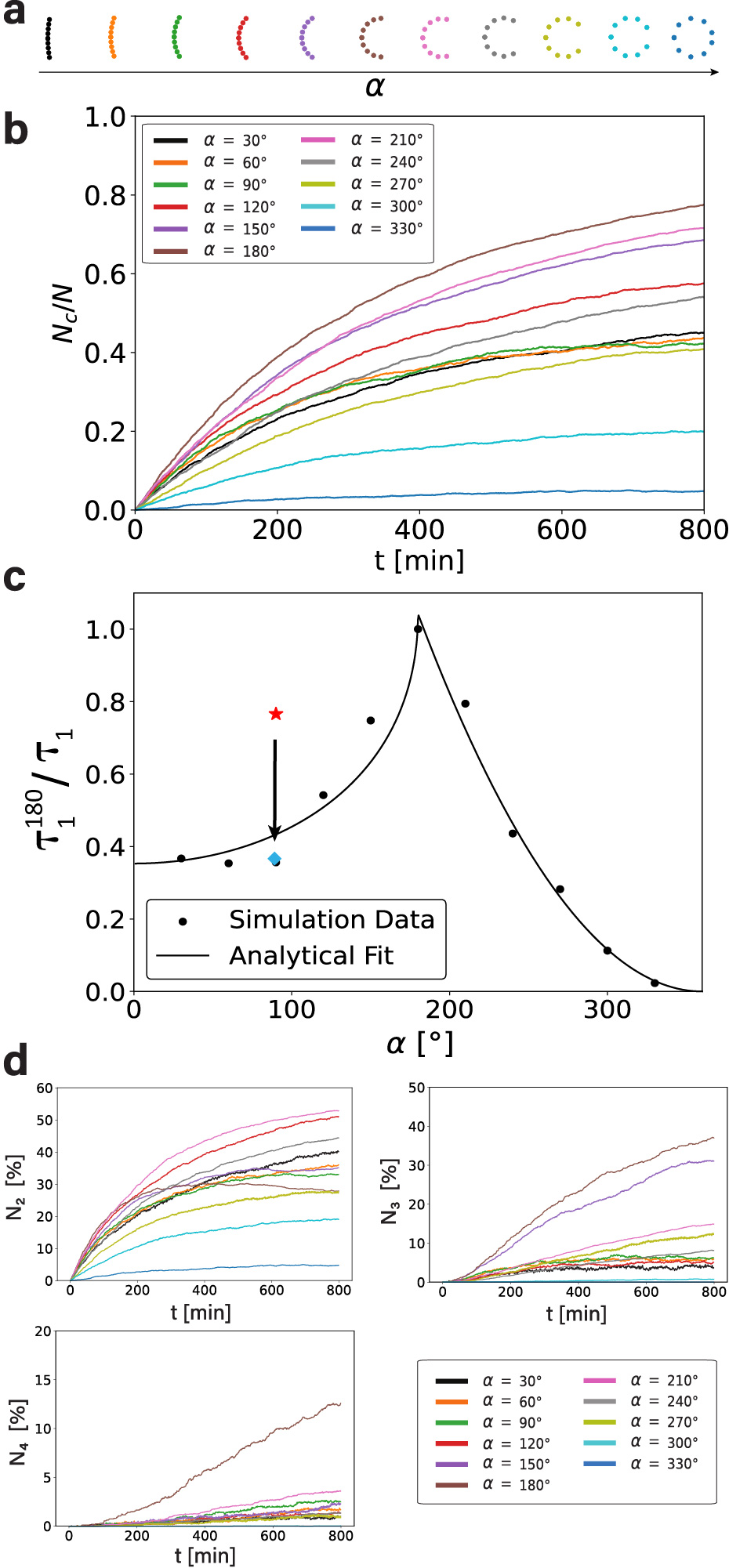}
\caption{\textbf{Simulation results on the influence of the opening angle $\alpha$ on the clustering behaviour} (a) Bent rods with increasing opening angle from $\ang{30}$ to $\ang{330}$ used in the simulations shown in (b). (b) Evolution of $N_c/N$ over time for active crescents with fixed cross-sectional length ($L = 10$ \si{\micro\meter}) but different opening angles for $N = 66$ particles each. (c) Inverse fit parameter $1/\tau_1$ as a function of the opening angle $\alpha$, normalized to the value of $\tau_1$ for $\alpha =\ang{180}$. The fit is found from a simple model describing cluster creation and decay, see the main text for details. The fit shown here is $\tau_1^{180}/\tau_1 = 2.357 (1.149 - 1/\sqrt{d(\alpha)^2 + 1}) \Theta(\pi - \alpha) + 0.106 P_{\rm clu} \Theta(\alpha - \pi)$, with $\Theta$ the Heaviside step function. The red data point (star) represents the normalised inverse fit parameter for 90\textdegree{} crescents $\left(\tau_1^{180}/\tau_1^{90}\right)$ found in experiments before rescaling, the blue data point (rectangle) the value after rescaling (see main text). (d) Evolution over time of the number of particles in a two-particle-cluster (top-left), a three-particle-cluster (top-right) and a four-particle-cluster (bottom-left), respectively.}
\label{Fig_5: shape dependence_simulations}
\end{figure*}
An even bigger shape change occurs if the particles move and interact with their convex side instead of their concave side, see Fig.~\ref{Fig_4: shape dependence}d, with effectively a collision interaction comparable to those of spheres. We experimentally achieve this by increasing the $H_2O_2$ concentration to 5\% and term these $c^+$ crescents, following Ref.~\cite{Wensink2014}.
While their average velocity at $1.27\pm0.23$~\si{\micro\meter\per\s} was slightly higher than that of the concave-side-leading particles, their efficiency to cluster at low density drops significantly (Fig.~\ref{Fig_4: shape dependence}e). The few pairs observed over the course of the experiment typically broke up by sliding past each other within $\sim$ 2 min (Supplementary Video 7). 
\\
\\
To study many different opening angles we turn again to simulations to avoid time-consuming 3D printing. We consider eleven different opening angles from 30 \textdegree{} to 330\textdegree{} in steps of $\Delta\alpha=30$\textdegree{}, keeping the cross-section fixed, see Fig.~\ref{Fig_5: shape dependence_simulations}a. For fixed arc-length and varying cross-section, the particles cluster at a similar speed, see SI. When again plotting $N_c/N$ we find that 180\textdegree{} crescents cluster most efficiently (cf. Fig.~\ref{Fig_5: shape dependence_simulations}b) with the smallest value for $\tau_1$. To quantify this behavior we present in Fig.~\ref{Fig_5: shape dependence_simulations}c the inverse fit parameter $1/\tau_1$ as a function of $\alpha$, normalized to the value found for $\alpha =180$\textdegree{}. The larger $1/\tau_1$, the faster the clustering. We find the largest value for 180\textdegree{} crescents, with $1/\tau_1$ being approximately symmetric about $\alpha = 180$\textdegree{} up to $90$\textdegree{}. However, increasing the opening angle further, $\tau_1$ increases sharply, while with decreasing opening angle the fit parameter reaches a plateau around $\alpha =90$\textdegree{}. 
\\
To rationalize the dependence of $\tau_1$ on $\alpha$ we assume two competing effects are relevant\footnote{This approach is similar to Ref.~\cite{Hoell2016}. However, a three-dimensional system of passive bent rods was considered there, which results in significantly different dynamics.}: 1) the probability $P_{\rm clu}$ to form a cluster and 2) the stability of the cluster. $P_{\rm clu}$ depends on the opening length $L_{\rm op}$, which is independent of $\alpha$ for $\alpha < 180$\textdegree{}. However, for $\alpha > 180$\textdegree{}, $L_{\rm op}$ and thus $P_{\rm clu}$, decrease with increasing $\alpha$. In the extreme example of a sphere, stable dimers cannot form. The probability of two particles colliding at an angle such that their cavities are facing each other, resulting in a dimer, is $P_{\rm clu}\sim (\alpha - 2\pi)^2 \sim 1/\tau_1$.
\\
However, once a cluster has formed, its stability depends on $\alpha$, too. For smaller $\alpha$, particles interlock less and cluster are less stable, e.g., in case of a collision with a free particle. As a simple proxy for stability we define the length $d(\alpha) = \sqrt{(L_{\rm op}/2)^2 + h^2}$ that one crescent in a dimer would have to move to leave the cluster. The smaller this length, the more unstable the cluster. This length scale yields a time scale $\mathcal{T} \sim d/v_0$ such that $\tau_1 \sim \mathcal{T}$ captures the $\alpha$-dependence of the parameter $\tau_1$. We find $ 1/\tau_1 \sim 1/d(\alpha)$. We find that combining these two effects results in a good agreement, see Fig.~\ref{Fig_5: shape dependence_simulations}c, especially considering the simplified arguments we used.
\\
Cluster formation is optimal for 180\textdegree{} crescents because  the probability for cluster formation is high, due to their relatively large opening length, and, once formed, clusters exhibit significant stability from the particles' strong curvature. In experiments we also find $1/\tau_1$ to be highest for 180\textdegree{} particles. However, $\tau_1^{180}/\tau_1^{90} =0.79$ for 90\textdegree{} crescents is significantly higher than the value found in simulations. We can rationalize this difference from the different propulsion speeds and sedimentation rates of the two particles. By rescaling $\tau_1^{180}/\tau_1^{90}$ with the measured propulsion velocities for the two shapes as well as with the inverse calculated sedimentation velocities (see SI and \cite{Tchen1954}), we obtain the corrected value of $\left(\tau_1^{180}/\tau_1^{90}\right)_{c}=0.38$ which is in perfect agreement with the value found in simulations, implying again that hydrodynamic interactions are negligible. 
\\
As a consequence of the shape-dependent competition between interlocking and breaking-up, higher-order cluster are also more stable for 180\textdegree{}. Three-particle-clusters are already most prevalent for 180\textdegree{} particles, but four-particle-clusters are effectively only observed for $\alpha = 180$\textdegree{} (see Fig. \ref{Fig_5: shape dependence_simulations}d). While the absence of higher order clusters in samples with higher opening angles can be explained by the increased difficulty to form these types of structures, for lower opening angles, the lack of larger clusters arises from the instability of two- and three- particle clusters when colliding with free particles. 
\\
\\
\textbf{Conclusions}\\
In this manuscript we investigated the clustering behavior of self-propelled particles with a shape that is continuously tunable along a single anisotropy dimension by exploiting a combination of 3D printing and simulations. Printing bent rods with different opening angles allowed connecting our results with the two major shapes thus far employed, spheres and rods. Surprisingly, stable interlocking into pairs occurs at arbitrarily low number concentrations of particles. Pair formation plays a critical role in nucleation, and the dynamics of the fraction of clustered particles can thus be captured with a notably simple formula which allows rescaling of the clustering dynamics into a single master curve. We find that particle shape influences both the assembly speed and cluster size distribution with an optimum for crescents with an 180\textdegree{} opening angle and can be described by a competition between interlocking and break-up.  
\\
Our experimental and numerical results provide the first detailed understanding on how a gradual shape change from a sphere to a rod affects the clustering behavior of active particles and uncover a simple yet general description for this complex effect. These key insights can be leveraged to rationally design lock-and-key type connections~\cite{sacanna_lock_2010} between active particles whose stability can be tuned through the depth of their entangling sites independent of their composition and surface chemistry. This strategy offers the unprecedented opportunity to design hierarchical assembly pathways towards functional materials through engineering the active particle shape.
The guaranteed assembly even at extremely low concentrations furthermore opens up possibilities in the time-delayed delivery of two-component drugs where a low concentration of the drug-carriers as well as the activation of reactants after assembly is important.
\\
\\
\textbf{Acknowledgements}
This work was supported by the Netherlands Organization for Scientific Research (NWO/OCW), as part of the Vidi scheme (S.R, L.A.H., L.G. and D.J.K.), and by the European Union via the ERC-CoGgrant HexaTissue (L.G.) We thank Rachel Doherty for support with 3D printing. Part of this work was performed using the ALICE compute resources provided by Leiden University.
\bibliographystyle{unsrt}
\bibliography{References}
\newpage
\section*{\textbf{Materials and Methods}}
\subsection*{Reagents}
\label{Reagents}
Unless otherwise noted, all chemicals were of analytical or reagent grade purity and were used as received from commercial sources. Fused silica substrates as well as the photoresist IP-Dip were acquired from Nanoscribe GmbH. Propylene glycol methylether acrylate (PGMEA, ReagentPlus $\geq$99.5\%) was purchased from Sigma Aldrich and isopropanol (IPA) was obtained from VWR chemicals. Hydrogen peroxide ($H_2O_2$, 35 wt\% solution in water, stabilized) was purchased from Acros Organics. Water was purified by means of a MilliQ system (resistivity \(\geq\) 18 M\(\Omega\).cm).
\\
\\
\subsection*{Particle Fabrication}
\label{Particle Fabrication}
All crescent-shaped particles were 3D printed on a Nanoscribe Photonic Professional GT microprinter equipped with a 63x oil-immersion objective (Zeiss, NA = 1.4). The exact printing routine has been described by us in detail in \cite{Doherty2020}. Particles were printed on a fused-silica substrate using IP-Dip as photoresist. Bent rods with an opening angle of 180\textdegree{} were designed with a thickness of 2 $\mu$m and a cross section length $L$ of 10 $\mu$m. For bent rods with an opening angle of 90\textdegree{} the thickness of the particles was reduced to 1.5 $\mu$m in order to keep the same size ratio between the thickness of the crescent legs and the concave opening. During development the print was submerged in PGMEA for 30 min followed by 2 min in IPA and then left to dry overnight at ambient conditions. Once dry, active crescents were coated with a 5 nm layer of catalyst (Pt/Pd 80:20) using a Cressington 208HR sputter-coater. To prevent excess Pt/Pd, the area around the print was protected with tape which could be removed later without damaging the printed structure. After sputter coating, each print was checked by default under the microscope to determine its quality. The substrate was placed onto a 200 $\mu$L MilliQ water droplet in the center of a small glass petri dish such that the printed crescents were in contact with the water. Particles could then easily be removed from the substrate by sonication for 2 min. The particle solution was collected in an Eppendorf tube and the procedure was repeated until approximately 90\% of the particles were released from the substrate. Particles were subsequently concentrated by centrifugation and removal of the supernatant. Such particle stock solutions could be stored in the fridge for several months.     
\\
\\
\subsection*{Imaging and Data Analysis}
\label{Imagingand Data Analysis}
Activity induced cluster formation of crescent-shaped particles was observed using a Nikon Eclipse Ti-E bright-field light microscope with a Plan Apo $\lambda$ 20x long working distance objective (NA = 0.75). Pt-coated colloidal particles were suspended in freshly prepared 1\% hydrogen peroxide aqueous solution in the case of concave-side-leading crescents or in 5\% hydrogen peroxide for convex-side-leading crescents. Control experiments with Brownian crescents were performed in MilliQ water. This colloidal solution was then placed into a sample holder ($\varnothing$ = 8 mm) using an untreated borosilicate glass coverslip (VWR, 25 mm, No. 1) as substrate. The cluster distribution in the sample was recorded after different time steps. We hereby defined a cluster as an assembly of at least two particles that remained stable for minimum 5s. The observation area (4.5 mm x 4.5 mm) was restricted to the central area of the sample holder and 7x7 images (2048x2048 px) with an overlap of 1\% were taken to cover this entire observation region. The time needed to take these 49 single images including corresponding 5s videos for each field of view resulted in an average value and error for the determination of the time-point. Short videos were only taken for samples with lower particle densities where the total observation time exceeded 100 min. Due to the higher fuel concentration needed to obtain $c^+$ crescents, the particle density had to be kept relatively low and the observation times short. Otherwise the formation of bubbles hindered the adequate conduction of our experiments. All measurements were taken in the dark.
\\
Single crescents as well as crescent cluster species were counted for every time step. From short (5s) videos, the average amount of stuck single crescents was determined. Stuck crescents were later removed from the previously determined number of single crescents. Average particle velocities were captured from 30s long videos with a frame rate of 20 fps. Particles were tracked in each frame, applying the Canny Edge detection algorithm to generate a mask out of which the particle center of mass was obtained. Individual crescent-velocities $v$ were determined by fitting the short-time regime ($\Delta t \ll \tau_r$) of the mean squared displacemnt (MSD) of different crescents with $\Delta r^2 = 4\mathcal{D}\Delta t + v^2 \Delta t^2$. The second fit parameter is the diffusion coefficient $\mathcal{D}$ and $\tau_r$ was chosen as the rotational diffusion time for a sphere. We here use the aforementioned fit-equation, originally developed for spherical active particles (Howse 2007, Bechinger 2016), as an approximation for our crescent-shaped particles despite them not being spherical.
\\
\\
\subsection*{Simulations}
\label{Simulations}
To simulate the experimental system we follow the model of Wensink et al.~\cite{Wensink2012}. We simulate $N$ self-propelling particles in two dimensions that move with a velocity $v_0$. We employed the same particle dimensions and velocity in the simulations as in experiments, see SI for a mapping between computational and physical units. The dynamics is assumed to be overdamped and particles interact only by steric repuslion with each other, i.e., there is no hydrodynamic interaction. To build a particle it is discretized into $i = 1, ..., k$ equidistant spherical segments, each with diameter $d$. A particle $\rho$ has an orientation $\bm{u}_\rho$ and a position $\bm{r}_\rho$. The position of segment $i$ with respect to the position of the center of mass $\bm{r}_\rho$ is denoted by $\bm{e}^i_\rho$. 
The pair potential of two particles $\rho$ and $\delta$ is given by $U_{\rho \delta} = k^{-2} \sum_{i,j = 1}^k u\left(r_{\rho \delta}^{i j}/d\right)$. Here, $u(x) = u_0 \exp\left(-x\right)/x^2$ is a short-range potential that is repulsive if $u_0 > 0$.  This results in effectively hard particles. $r_{\rho \delta}^{i j} = \left| \bm{r}_\rho - \bm{r}_\delta + \bm{e}^i_\rho - \bm{e}^j_\delta \right|$ is the distance of two segments of the two different particles. The equations of motion, found, from balancing forces and torques due to activity and steric repulsion, for a particle $\rho$ are given by~\cite{Wensink2012}:
\begin{subequations}
\begin{align}
f_t \partial_t \bm{r}_\rho &= F_{\rm a} \bm{u}_\rho - \bm{\nabla}_{\bm{r}_\rho} U + \xi \, \\
f_r \partial_t \varphi_\rho &= - \bm{\nabla}_{\varphi_{\rho}} U + \eta \;\,
\end{align}
\label{eq:EOM_Simulation}
\end{subequations}
where $f_t$ and $f_r$ are translational and rotational friction, respectively, $\bm{u}_\rho = \left\{\sin \varphi_\rho, \cos \varphi_\rho\right\}$, $v_0 = F_{\rm a}/f_t$, and $U = \sum_{\rho, \delta (\rho \neq \delta)} U_{\rho \delta}/2$. $\xi$ and $\eta$ are translational and rotational Brownian noise, respectively. We simulate the particles in a square system with periodic boundary conditions, see the SI for details.
\\
\\
\subsection*{Analytical model}
\label{Analytical}
Because of the dominant role that pair formation plays in the nucleation of the clustering process, we model the clustering dynamics using a simple balls-into-bins model. For this we assume an initial number of $N$ particles of cross-sectional length $L$ which move with a speed $v_0$ in a square system of area $D^2$. We divide the system into square boxes of width $S = L$ such that there are $(D/L)^2$ boxes in total. In the first step we randomly assign a box to each of the $N$ particles. If two (three, etc) particles are assigned the same box this is counted as a two (three, etc) particle-cluster. In the next step, particles that are part of a cluster remain in their box while the remaining free particles are randomly assigned to a box again. If, for example, a particle is assigned a box that was already occupied by a two cluster this results in a three cluster. For more details see SI. The results of the simulations for the clustering dynamics over time, i.e., the $N_c/N$ curve, are shown in Fig. \ref{Fig_2: time dependence}g, and look qualitatively similar to those of the experiments and simulations. 
For simplicity we now consider only two-clusters forming since these are dominating the experimental dynamics, at least at small times or low concentrations. From simple probabilistic arguments we find then that this process can be approximated as a Poisson process with time-dependent rate, see the SI for details. That is, the average cluster size is given by $\langle N_c(t) \rangle = 2 r(t) t$, where $r(t)$ is the rate at which clusters form at a given (dimensionless) time $t$, and the factor of two is due to two particles being in a cluster. $N_c(t)$ is the number of particles that are in a cluster at a given time. We call $N_f$ the number of free particles, such that $N = N_c(t) + N_f(t)$. To determine the rate we need to consider the probability that, for $N_f$ balls and $B$ bins, two balls are assigned the same bin. This probability is given by the ``birthday-paradox'' probability
\begin{equation}
\label{eq:ApproxRate}
r(N_f,B) \approx 1 - \left(1-\frac{N_f}{2B}\right)^{N_f-1} \;,
\end{equation}
which is a good approximation of the exact expression (see SI) for $B/N_f \gg 1$.
If the number of balls one draws from were constant, $N_f(t) = N$, then the rate $r(N,B)$ is constant in time and the number of particles in a cluster is linearly increasing in time, $N_c(t) \sim t$. In the experiments this is approximately the case if $N_c/N \ll 1$ since, in this case, $N_f(t) \approx N$. See, e.g., Fig. \ref{Fig_2: time dependence}f. However, if the number of free particles decreases significantly over time, the rate is time-dependent. Assuming this decrease is linear, $N_f(t) = N - 2 r(t) t$, we find from Eq. \eqref{eq:ApproxRate} the implicit equation
\begin{equation}
r(t) \approx 1 - \left(1-\frac{N - 2 r(t) t}{2B}\right)^{N - 2 r(t) t - 1} \;. 
\end{equation}
for the rate $r(t)$. For $B \gg N \gg 1$, i.e. small $r(t)$, we can Taylor expand this expression and solve for $r(t)$. We refer to the SI for further details on the derivation. We find the simple relation:
\begin{equation}
\frac{\langle N_c(t) \rangle}{N} = \frac{t}{B/N + t} \;.
\label{eq:FitFuncFromBiB_v2}
\end{equation}
Note that for small $t$ the increase is linear, $\langle N_c \rangle/N \sim (N/B) t$, recovering the case where the rate is essentially time-independent. For large $t$ all particles are in a cluster, $\langle N_c \rangle/N \to 1$. Thus, there is only one free parameter in Eq.~\eqref{eq:FitFuncFromBiB_v2}, namely $B/N$, essentially system size over particle number, i.e., the inverse density. This parameter is the time needed for half the particles to be part of a cluster, $\langle N_c(t = B/N) \rangle = N/2$. Eq.~\eqref{eq:FitFuncFromBiB_v2}, with $B/N$ as a free parameter, can be used as a fit for the experiments and the balls-into-bins simulations. As can be seen in Fig. \ref{Fig_2: time dependence}f,~g it matches well. 
\newpage
\section*{\textbf{Designing highly efficient lock-and-key interactions in anisotropic active particles - Supplementary Information}}

\section*{3D printing of bent rods}

\begin{figure}[h!]
\centering
\includegraphics[width=0.7\textwidth]{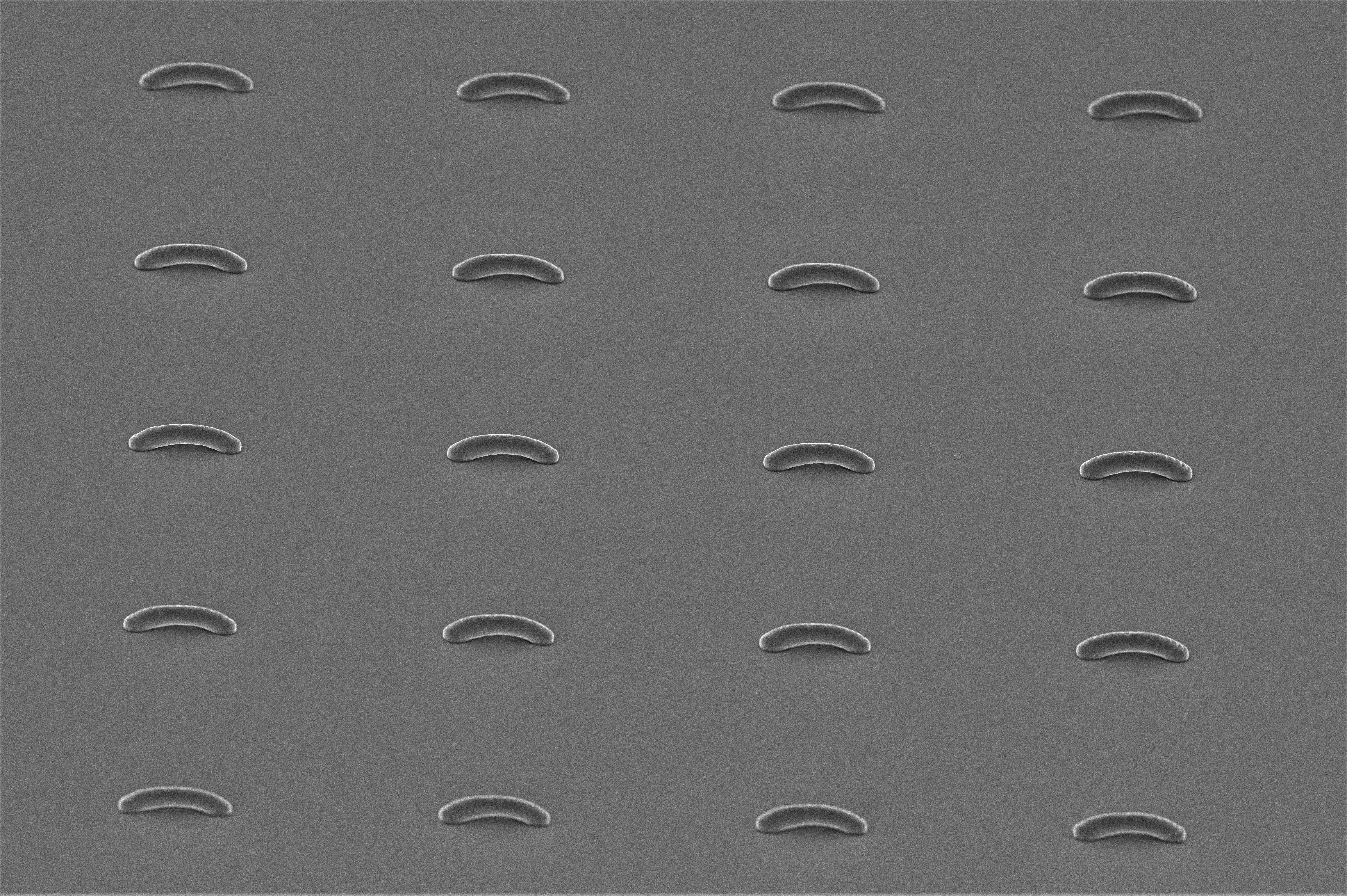}
\caption{Scanning electron microscopy image of an array of 3D printed bent rods with an opening angle of $\alpha=\ang{90}$ (90\textdegree{} crescents).}
\label{fig:SEM image}
\end{figure}
\section*{Self-organization of active concave-side leading 180\textdegree{} crescents}

\begin{figure}[h!]
\centering
\includegraphics[width=0.5\textwidth]{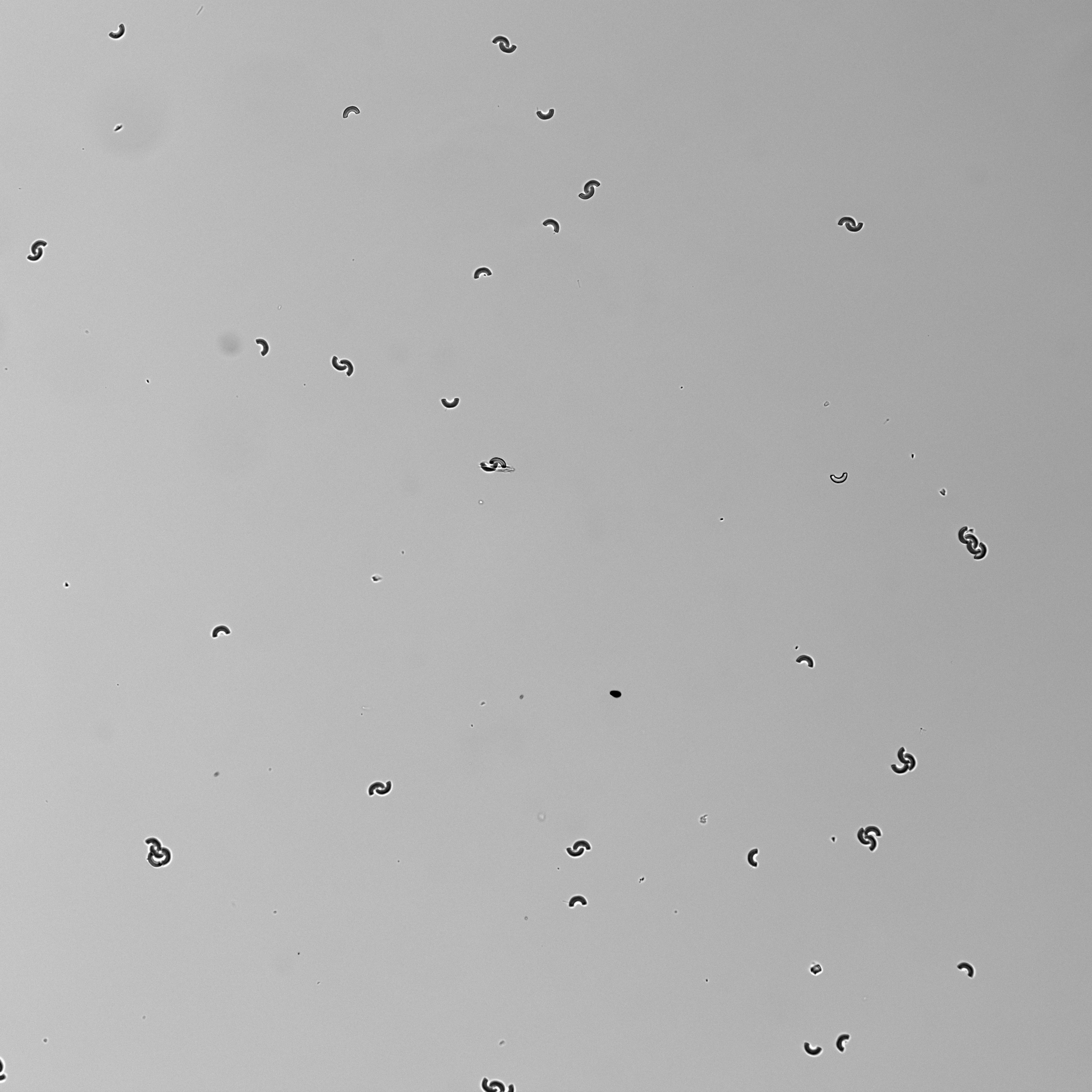}
\caption{Full field of view taken 90 min after mixing the particles and the fuel solution ($\phi_s=0.263\%$). A detail of this field of view is shown in Fig. 2a. Contrast and brightness were increased to improve the clarity of the image.}
\label{fig:full field of view active}
\end{figure}
\newpage
\section*{30 sec trajectories of active concave-side leading 180\textdegree{} crescents}

\begin{figure}[h!]
\centering
\includegraphics[width=0.8\textwidth]{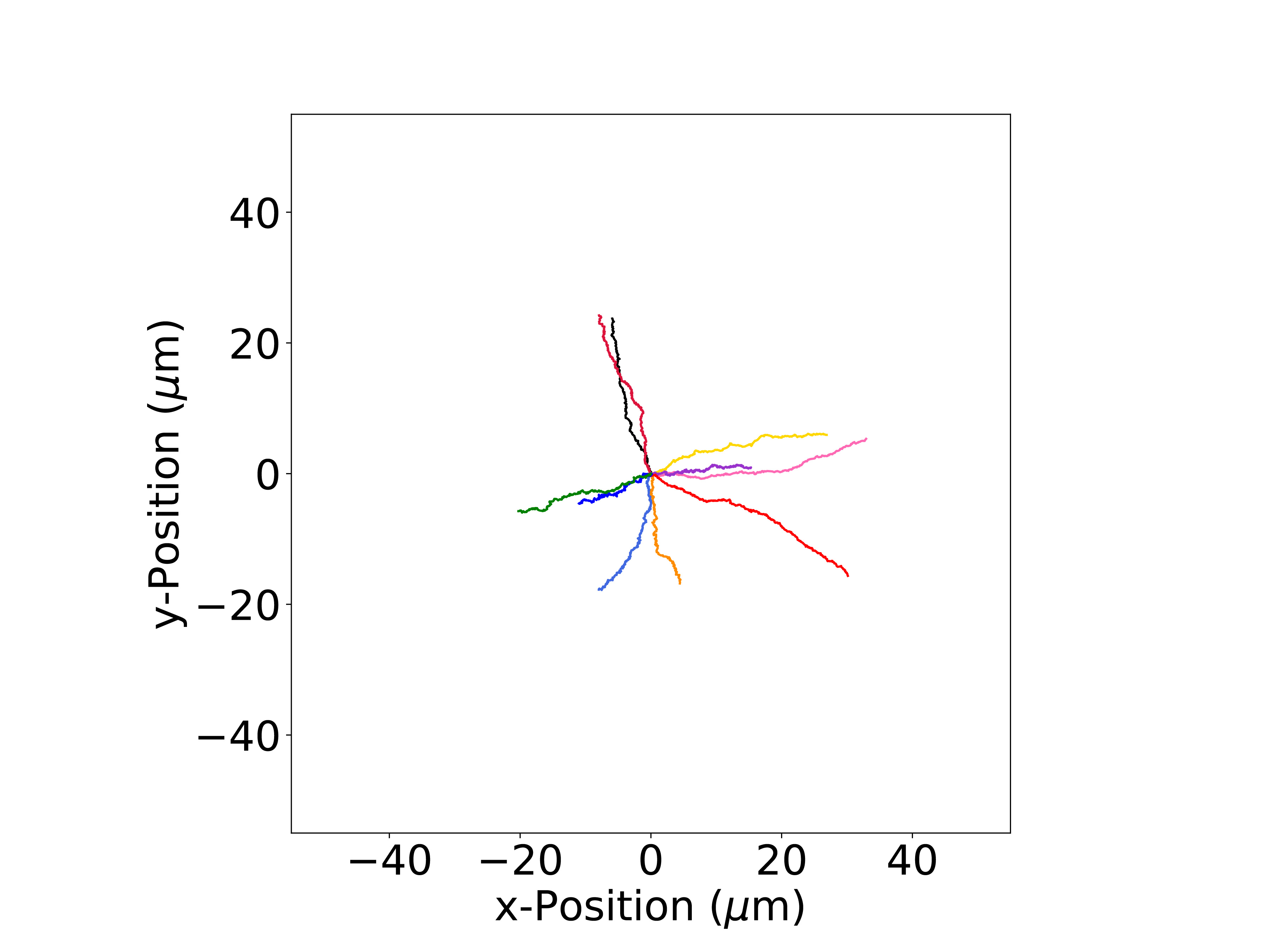}
\caption{30 sec trajectories for concave-side leading 180\textdegree{} crescents. Due to their size and shape, the motion of these active crescents shows long persistance length.}
\label{fig:trajectories}
\end{figure}
\newpage
\section*{Passive 180\textdegree{} crescent-shaped particles}

\begin{figure}[h!]
\centering
\includegraphics[width=0.7\textwidth]{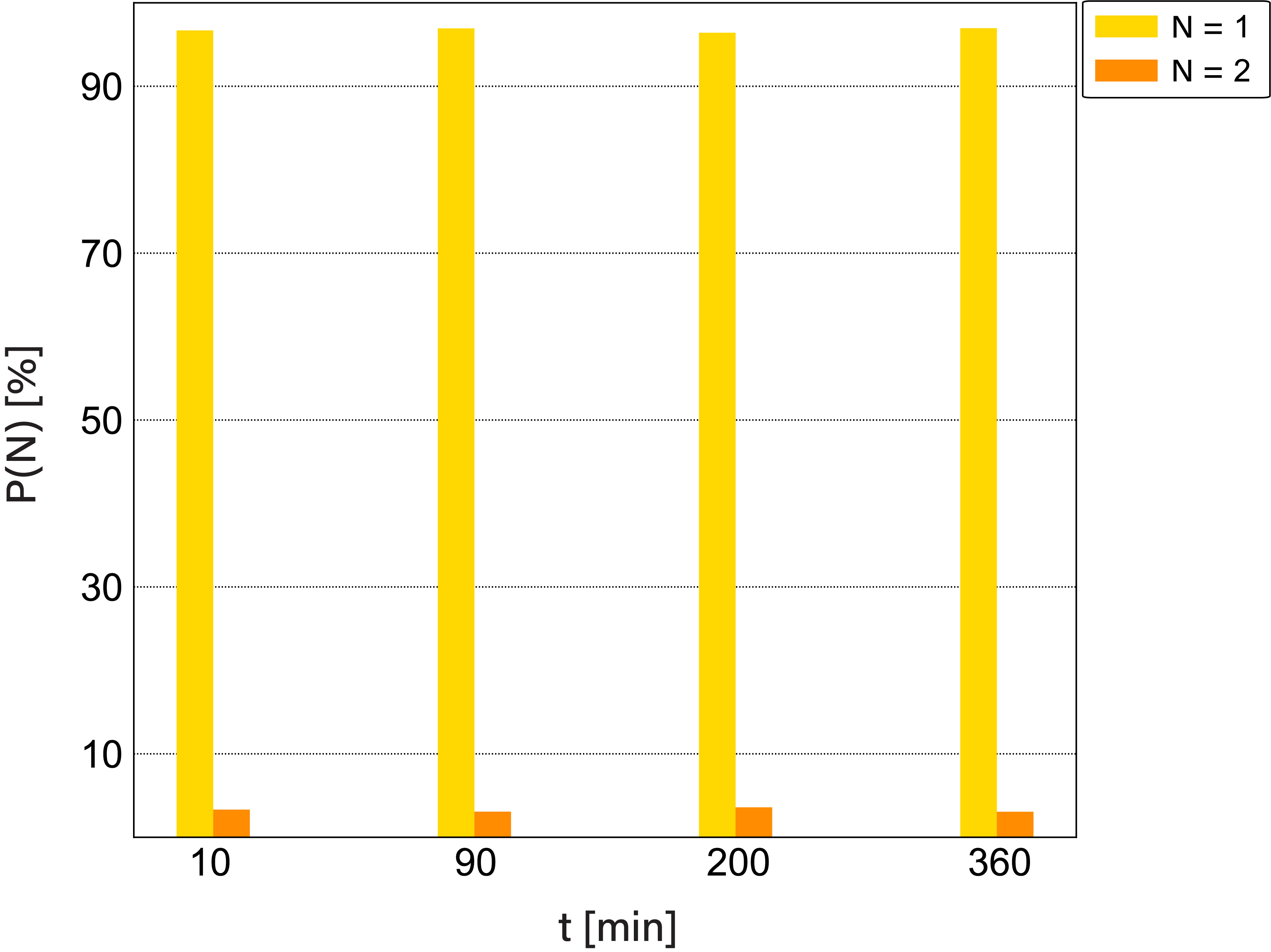}
\caption{Cluster distribution over time for a sample of passive 180\textdegree{} crescents suspended in water with a surface area fraction of ($\phi_s=0.237\%$) after 90 min. This corresponds to a particle density of 94.4 crescents/mm2 Passive crescents are not coated with a 5nm Pt-layer.}
\label{fig:cluster size distribution for passive}
\end{figure}

\begin{figure}[h!]
\centering
\includegraphics[width=0.5\textwidth]{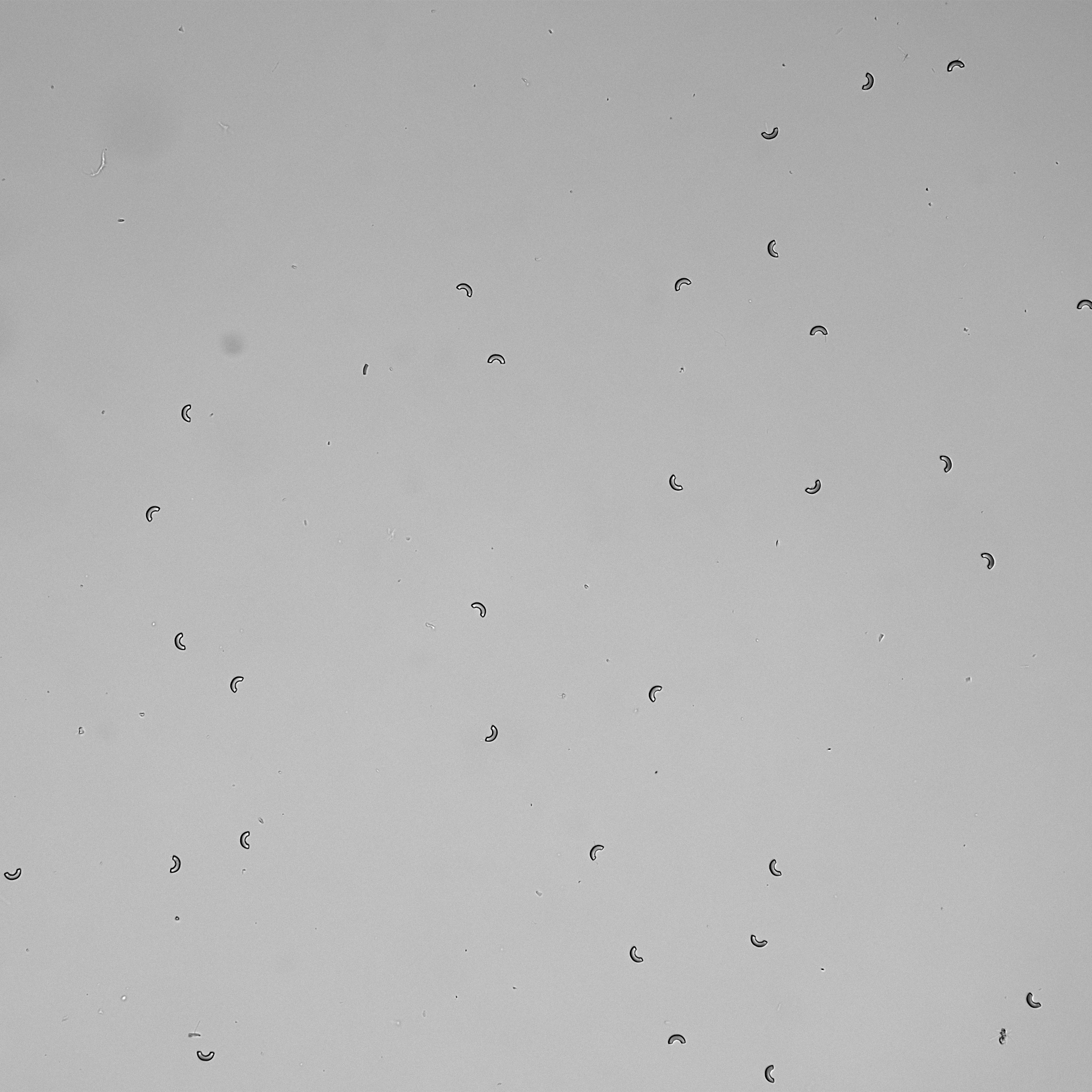}
\caption{Full field of view taken 360 min after suspending the particles in milliQ water ($\phi_s=0.237\%$). Contrast and brightness were increased to improve the clarity of the image.}
\label{fig:full field of view passive}
\end{figure}
\newpage
\section*{Cluster distribution and density profile for convex-side leading 180\textdegree{} crescents}
A decrease in particle density over time for convex-leading 180\textdegree{} crescents can be explained by the fact that these crescents sediment faster than their concave-leading counterparts. Therefore sedimenting crescents stop compensating stuck crescents earlier in time. Since stuck single crescents are not taken into account when determining the total number of  crescents, the density of particles decreases once all crescents have sedimented. We are aware that this means that clusters are overrated when the particles density decreases. However, this does not have a major impact on our results. 

\begin{figure}[h!]
\centering
\includegraphics[width=1.0\textwidth]{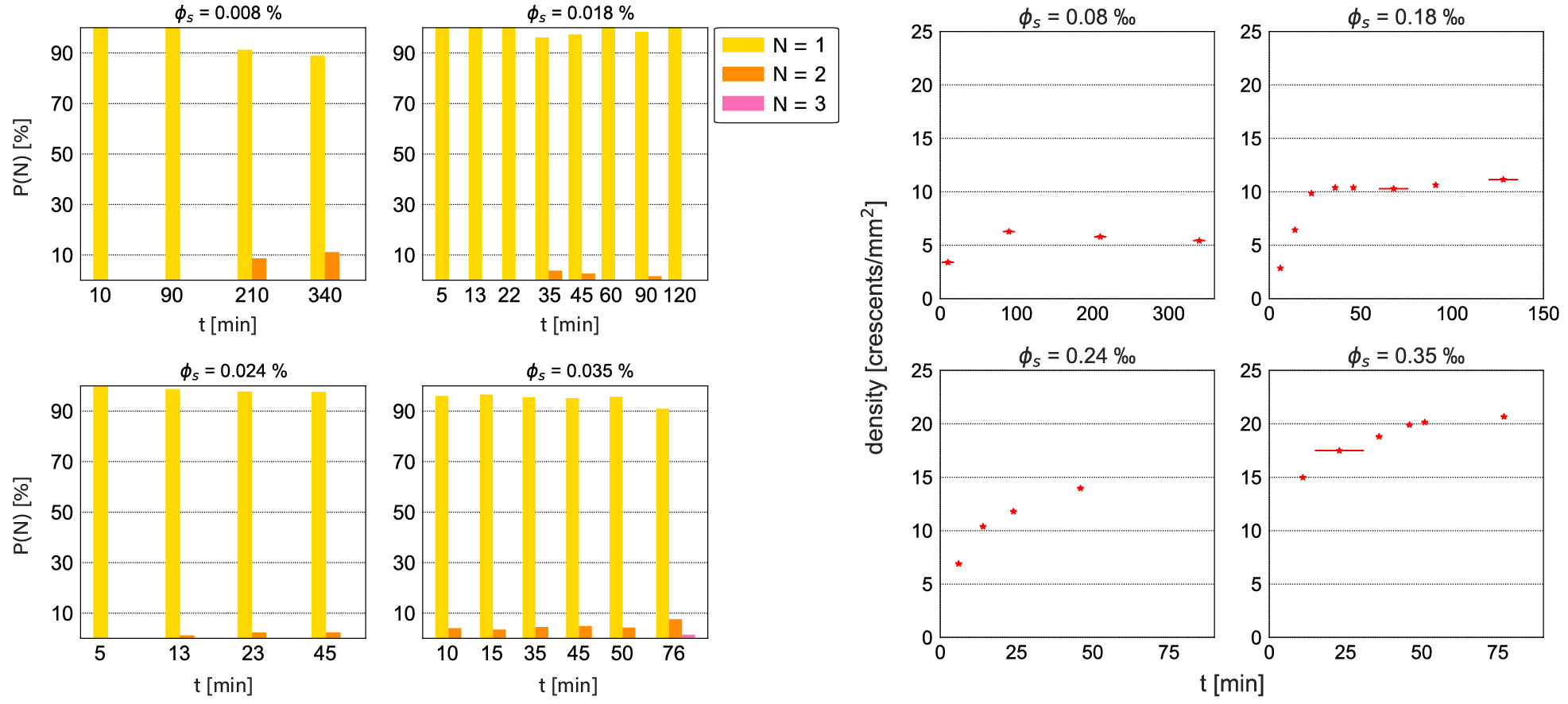}
\caption{Cluster size distributions (left) and corresponding density profiles over time (right) for samples of $c^+$ crescents with different surface area fractions $\phi_s$. Samples correspond to the results shown in Figure 4b of the main text. Stuck crescents are taken into account in the density profiles.}
\label{fig:cluster size distribution for c+ crescents}
\end{figure}

\begin{figure}[h!]
\centering
\includegraphics[width=0.9\textwidth]{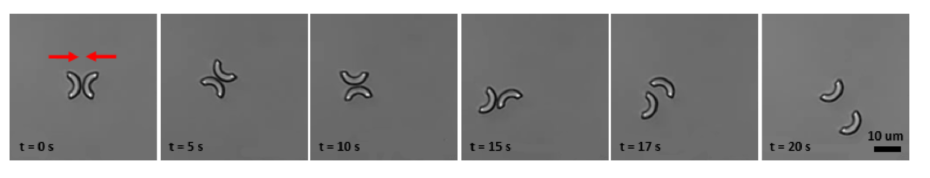}
\caption{Pairs of convex-side leading crescents are not stable beyond maximum 2 min.}
\label{fig:180deg c+ pair}
\end{figure}
\newpage
\section*{Density profiles for concave-side leading crescents}

\begin{figure}[h!]
\centering
\includegraphics[width=1.0\textwidth]{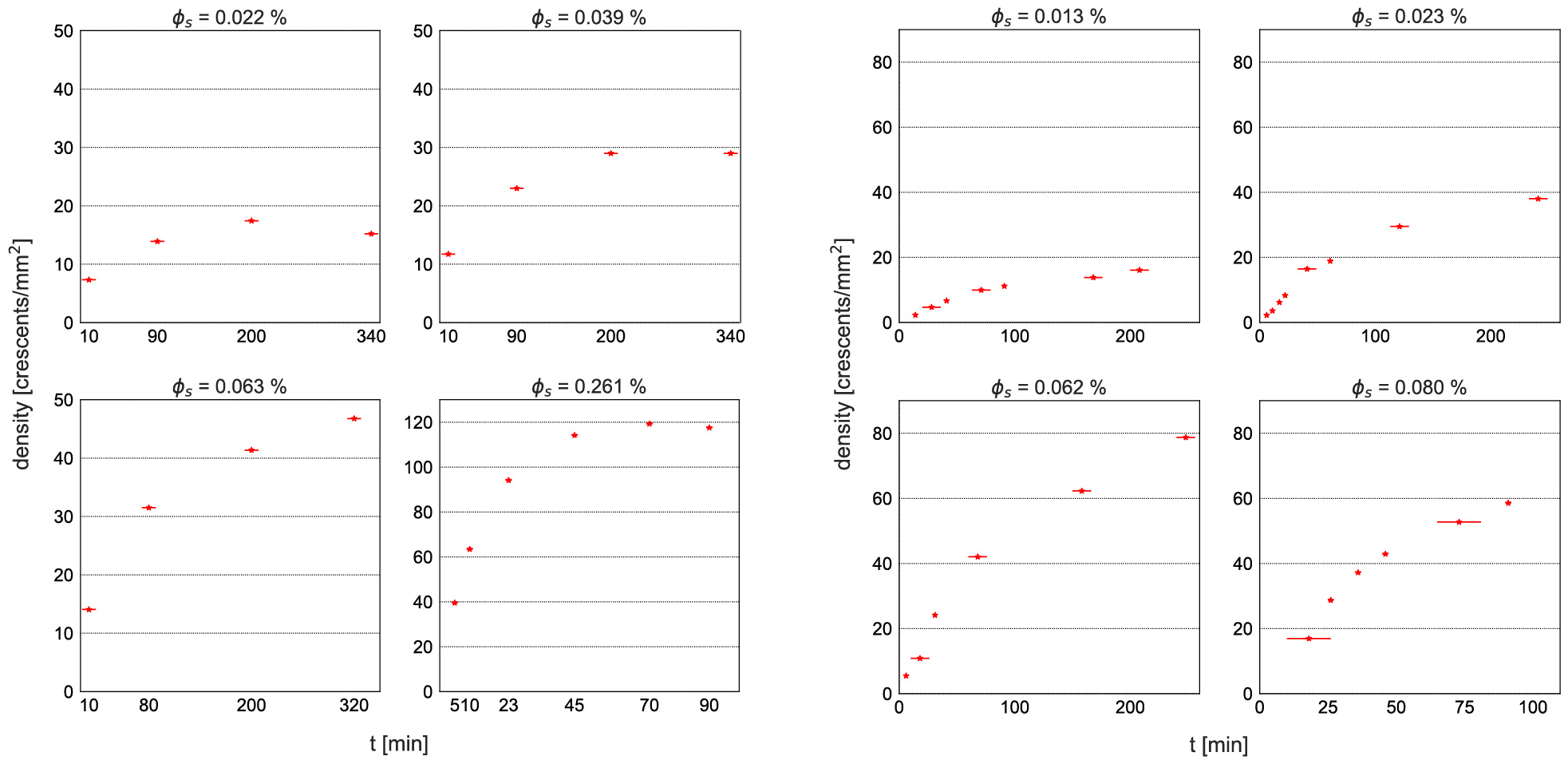}
\caption{Density profiles for all 180\textdegree{} (left) and 90\textdegree{} (right) concave-leading crescent-samples mentioned in our work. In both cases the cross-section of the particles is 10 \si{\micro\meter}. Stuck crescents are taken into account in the density profiles.}
\label{fig:concentration profiles for c- crescents}
\end{figure}

\section*{Dynamic simulations}
We present here some additional details on the simulations we performed, expanding on the paragraph in the methods section at the end of the main text. As described there a single particle is from $k$ equidistant spherical segments, each with an effective diameter $d$. The diameter is set by the exponential short-range pair potential we use to model the particles as hard spheres. According to the specific opening angle $\alpha$ and radius $R$ of a given particle we distribute the $m$ spheres along a circle segment of radius $R$. Each particles moves with a velocity $v_0$.
A number $N$ particles is initialized in a system of size $D \times D$ with random orientation and random position. We use periodic boundary conditions and choose the system to be the same size as the one used in the experiments (see below). Varying the system size by as much as a factor of 9, we found that the periodicity of the boundary is irrelevant for the clustering dynamics. The dynamic of each particle is given by the equations in Materials and Methods (Eq.~(2)). In the simulations we have defined a particles to be in a cluster with another particles if it is closer than $3R$ for 100 iterations.
Unless otherwise noted, each of the results presented is found by averaging over 100 independent runs.
We use the following parameters for the simulation: $D = 1137$, $v_0 = 0.04$, $d = 0.5$, $u_0 = 1$, $f_t = 1$, $f_r = 1$, $m = 9$, $\xi = 0$, $\eta = 0$. For the 180\textdegree{}-particles we have used $R = 1$. Hence, if a single such particle is initialized in the system of size $D \times D$, the area surface fraction is given by $\phi_s \approx 0.000012$\%. This can be used to easily convert number of particles in the simulations into area surface fraction. For other opening angles we have adapted $R$ according to if arc-length or cross-section is fixed. 
To map the quantities of the simulations into physical units we used the following experimental values: $v_0 = 1 \mu m/s$ and $R = 4 \mu m$ for a 180\textdegree{} particle. We thus find convert lengths and times from simulation into physical units as follows: $1 \text{ length} = 4 \mu m$, and $1 \text{ time step} = 0.16 s$.

\begin{figure}[h!]
\centering
\includegraphics[width=0.2\textwidth]{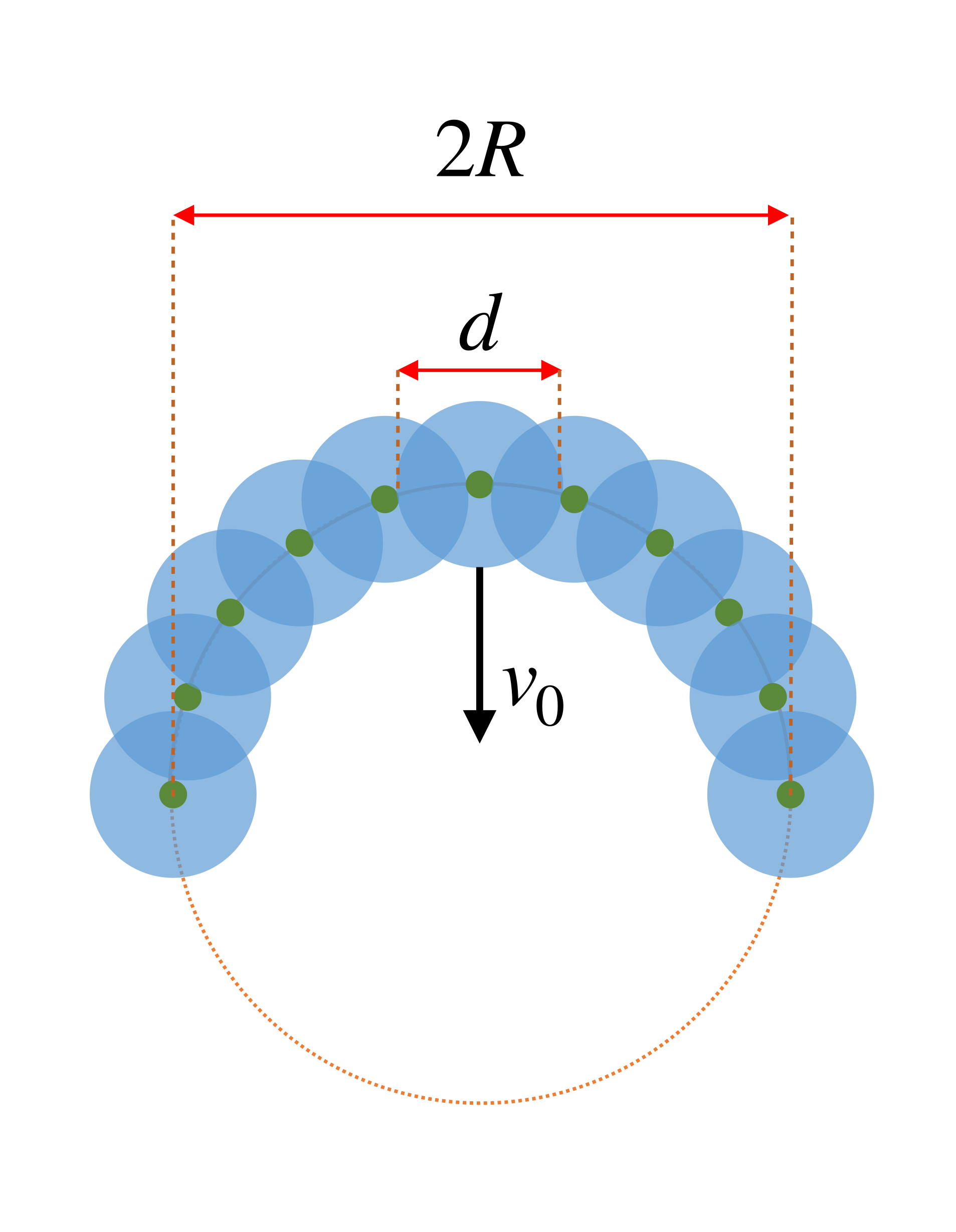}
\caption{\textbf{Sketch of particle used in simulation.} A particle of Radius $R$ and opening angle $\alpha$ is constructed by arranging discs of radius $d$ on a spherical segment. The particle moves with a velocity $v_0$.}
\label{fig:Sketch_Part_Sim}
\end{figure}

As mentioned above, we set the translational and rotational noise to zero in the simulations presented in the main text. We have compared simulations including the rotational and translational noises with values measured in the experiments in the experimental setup and compared simulations with and without noise. Including the noise did not change the behavior significantly. Thus, in the spirit of using a minimal model, we do not include it in the simulations presented below.

\newpage
\section*{Balls-into-bins model}

It is possible to model the clustering of the active particles as a simple probabilistic process for which we can also find some analytical expressions and results. In the following, we first run a simulation of a balls-into-bins model that yields results very similar to the experiments and dynamics simulations of the active particles. Second, we can find an analytical expression for the expected number of particles in a cluster for this balls-into-bins model. However, the analytical expression is very complicated and not very insightful. Thus, in the third part, we show that at short times there is a linear regime that corresponds to a Poisson process. The plateauing that is observed at longer times can then be described in this framework as a time-dependent rate that decreases over time.

\begin{figure*}
\centering
\includegraphics[width=\textwidth]{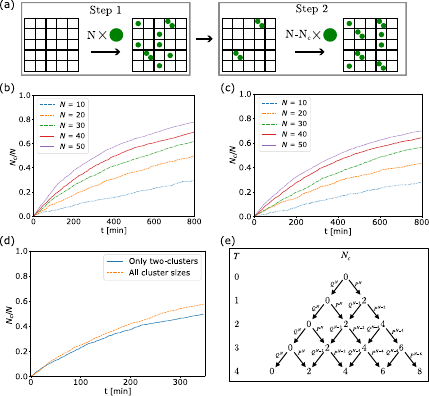}
\caption{\textbf{Balls-into-bins model.} (a) Sketch of the two steps in the balls-into-bins model. In Step 1 we assign each of the free particles (initially $N$ particles) randomly to a box. In step 2 we remove all particles that were in a box by themselves and assign these again to a new box. (b) We present the results from the simulation of the balls-into-bins model for different initial values $N$. We used the parameters $D = 1200$, $S = 5$, $v_0 = 0.04$. In (c) we present the results of the dynamic simulations for the same number of active particles. In (d) we compare the results of the balls-into-bins simulation for moderately high concentrations ($N = 60$) when taking only two clusters into account with the simulation where all clusters are taken into account. The curves deviate for larger times but the deviation is relatively small. (e) Presents the first four steps of the tree diagram used for the derivation of the exact expression for the average number of particles in a cluster $\langle N_c \rangle$ in Eq. \eqref{eq:ExactBiB}.}
\label{fig:BiB}
\end{figure*}

\textbf{Simulation of balls-into-bins model} We have initially $N$ active particles of cross-sectional length $L$ which move with a speed $v_0$ in a square system of width and height $D$. The dynamics of the active particles are mapped onto a balls-into-bins model as follows. The system is divided into boxes of size $S$; thus there are a total of $B = (D/S)^2$ boxes. The size $S$ of a box is determined by the size of the particle and we set $S \approx L$, assuming for simplicity that that if two active particles are in contact they will form a cluster.
In the first step of the simulation we put each of the $N$ particles randomly into one of the boxes. If two (three, ...) balls are assigned to the same box, we count this as a two- (three-, ...)cluster. If a ball is in a box by itself, it is counted as a free particle. In the following step of the simulation the above procedure is repeated but only all the \textit{free} particles are assigned a new box. That is, if in the previous step particles formed a cluster, they are kept in this box and are not randomly assigned a box again. See Fig.~\ref{fig:BiB}a for a sketch. This models the purely rotational, and absence of translational, motion of clusters that we observe in the experiments. This procedure is repeated $M$ times in total. If, at some point, a particle is assigned to a box that is already occupied by two (three, ...) particles, this is count this as a three (four, ...) cluster. Note that in this model a cluster, once formed, will never decay. This corresponds to the zero-noise limit and neglects potential collisions between free particles and clusters that can result in the destruction of the clusters, resulting in a finite lifetime. However, these are second-order effect, and, as a first approximation, these can be ignored for the sake of simplicity of this model.

We can convert the steps in the simulations into physical time steps as follows. We approximate that each time step in the simulation occurs after sufficient time has passed for an active particle to move to another box, i.e. each step of the simulation corresponds to a physical time step $\Delta t = S/v_0$. Thus, the total time the simulations run is given by $t_\text{total} = M S / v_0$. Some results of the simulations are shown in Fig. \ref{fig:BiB}b. It is easily seen that the shape of the resulting curves is very similar to the ones found in experiment and dynamical simulations, see Fig. \ref{fig:BiB}c. A quantitative agreement can be found by fine-tuning the parameters $N$, $S$, and $L$.

\textbf{Analytical expression for the simulation} To derive a general analytical expression for this balls-into-bins model we consider only two-clusters forming. That is, in the following we ignore the possibility of higher-order cluster forming. This is justified by the observation that two-clusters are dominant (at least at short times and/or low concentrations). In principle, it is straightforward to allow for higher-order clusters in following derivation. However, for the sake of simplicity we opt to consider the simplest case, which we find to agree well with the simulations. That is because the probability of a three-cluster forming in the simulations for $N \ll B$ is small. Indeed, running the above simulation under the same assumption (only two-clusters form) results in a curve very similar to the ones where all cluster sizes are taken into account, see Fig. \ref{fig:BiB}d. Furthermore, we assume that at each time-step of the simulations at most one cluster can form. Again, this is justified by the probability of a cluster forming for  $N \ll B$ being very small.

The main difficulty of the analytical derivation is due the probability of a cluster forming is potentially changing at each time step of the simulations because, once a cluster has formed, the number of balls one draws from decreases. Namely, the probability that, for $N_f$ balls (free particles) and $B$ bins, two balls end up in the same bin is given by the ``birthday paradox'' probability
\begin{equation}
P^{N_f} \equiv P(N_f,B) = 1 - Q(N_f,B) = 1 - \frac{B!}{(B-N_f)! B^{N_f}} \;.
\end{equation}
As the number of free particle decreases over time, this probability is dependent on the time step as well. Thus, for an initial state of $N$ balls we have one two-cluster after the first step with probability $P^N$ and no cluster with probability $Q^N$. We can then draw a tree diagram for the following steps, see in Fig. \ref{fig:BiB}e. For every two-cluster forming the number of free particles decreases by two, that is the probabilities become $P^{N-2}$ etc. From this one can compute the expected number of particles in a cluster after $T$ steps. From the tree diagram we find the following expression for the number of particles in a cluster after $T$ steps:
\begin{equation}
\label{eq:ExactBiB}
N_c(T,N) = \sum_{n = 2, n \in 2 \mathbb{Z}}^{\text{min}(2 T, N)} \left\{ n \left[ \prod_{k = 0}^{\frac{n}{2}-1} P(N - 2 k) \right] \sum_{m_0 + m_1 + ... = T - \frac{n}{2}} \prod_{j = 0}^\frac{n}{2} Q^{2 m_j}(N - 2 j)\right\} \;.
\end{equation}
While exact, this expression is computationally very expensive to compute for large $T$ as the number of terms grows exponentially with $T$. For this reason, we will derive a simplified expression below that is a good approximation for the cases of small density we consider experimentally.
Note that Eq. \eqref{eq:ExactBiB} simplifies considerably if the probabilities are assumed to be constant, which is true if $N_c \ll N$. In this case the complicated term (the sum in the last term) just reduces to a binomial coefficient and we find
\begin{equation}
\label{eq:ApproxBiB}
N_c(T,N) \approx  \sum_{n = 2, n \in 2 \mathbb{Z}}^{2 T} \left\{n \binom{T}{n/2} P(N)^{\frac{n}{2}} Q(N)^{T-\frac{n}{2}} \right\} \;.
\end{equation}
However, this expression is unbound, i.e., and $N_c \to \infty$ as $T \to \infty$ whereas Eq. \eqref{eq:ExactBiB} will plateau and $N_c \to N$ as $T \to \infty$.

\textbf{Approximation as Poisson process} We can find a simpler expression modelling the dynamics as a Poisson process. Again, we only consider two-clusters forming. To motivate the modelling as a Poisson process we first consider the case where $N$ is constant, i.e., the number of balls one assigns to a box is kept constant at each step. This corresponds to Eq. \eqref{eq:ApproxBiB}. As the number of bins is very large, $B \gg 1$, and the probability of a cluster forming is very small, $N \ll B$, the binomial distribution converges to the Poisson distribution. That is, the expected number of clusters is linearly growing in time with the proportionality constant being given by the rate $r$ at which clusters form. Here, the rate is given by the probability $P(N,B)$, i.e.,
\begin{equation}
\left\langle N_c \right\rangle =  2 r T = 2 P(N,B) T
\end{equation}
Here the factor of two is due to $P(N,B)$ being the probability of a two-cluster forming and, as there are two particles in each cluster, the number $N_c$ increases by two. For the low densities considered this is an excellent approximation of Eq. \eqref{eq:ApproxBiB}. 
A balls-into-bins simulation with a constant number of free particles results in such a linear relationship as well. Furthermore, for small $T$ it is a good approximation of Eq. \eqref{eq:ExactBiB} as well. 
However, for high concentrations or long times the constant-rate Poisson approximation breaks down as it will grow without bound because it assumes a constant rate, i.e. a constant number of free particles. Thus, we have to consider a time-dependent rate. First, we can simplify the probability as
\begin{equation}
r \equiv P(N_f,B) = 1 - \frac{B!}{(B-N_f)! B^{N_f}}  \approx 1 - \left(1- \frac{N_f}{2 B}\right)^{N_f-1}
\end{equation}
which for $B/N_f \gg 1$ is a very good approximation. Now, to include a time-dependent rate $r(T)$ we assume that the number of free particles is linearly decreasing in time with a rate set by $r(T)$, thus $N_f = N - 2 r T$. We then find
\begin{equation}
r = 1 - \left(1- \frac{N - 2 r T}{2 B}\right)^{N - 2 r T-1}
\end{equation}
which is an implicit expression for the rate $r$ that does not have an explicit solution. However, as the rate of clusters forming is small we can Taylor expand the right-hand side to lowest order in $r$. After some straightforward simplifications, assuming $N \gg 1$ and $B \gg N_f$, we find for $\left\langle N_c(T) \right\rangle =  2 r(T) T$ the simple expression
\begin{equation}
\frac{\langle N_c(T) \rangle}{N} \approx \frac{T}{\frac{B}{N} + T} \;.
\end{equation}
This expression contains only a single free parameter, $B/N$, essentially the inverse density.
Note that at small times this reduces to the linear function
\begin{equation}
\frac{\langle N_c(T) \rangle}{N} \approx \frac{T}{B/N}
\end{equation}
and thus we recover the Poisson process with constant rate. On the other hand, for large times we have $N_c(T) \to N$, and the function is bound.

\section{Histograms for 180\textdegree{}-particles from simulations}
\begin{figure*}[h!]
\includegraphics[width=\textwidth]{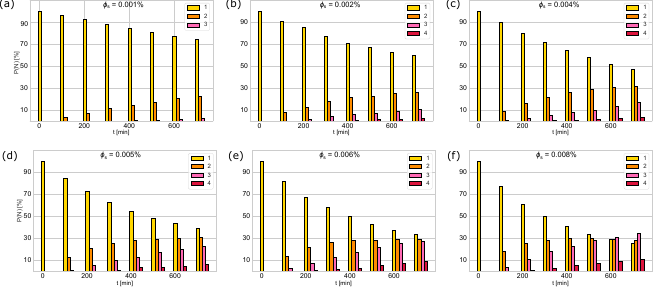}
\caption{\textbf{Histograms for 180\textdegree{}-particles from simulations}. We present the histograms of the cluster sizes over time for different concentrations. These are obtained from the dynamical simulations of the active particles with an opening angle of 180\textdegree{}.}
\label{fig:SI_Plots_180}
\end{figure*}

\section*{Time-dependent concentration}
As explained in the main text, we find that the concentration in the experiments is not constant but increases initially while particles are sedementing. We find that after a certain time the concentration plateaus and remains approximately constant afterwards. To investigate the large discrepancy of the clustering behavior we observe in experiments and simulations we consider a time-dependent concentration in the simulations as well. As described in the main text, if this is not taken into account we find that the active particles cluster significantly more and faster at a given concentration compared with the experiments. To this end, we measure experimentally how the concentration changes over time for different final concentrations. The results are presented in Fig. \ref{fig:concentration profiles for c- crescents}.

\begin{figure*}[h!]
\includegraphics[width=\textwidth]{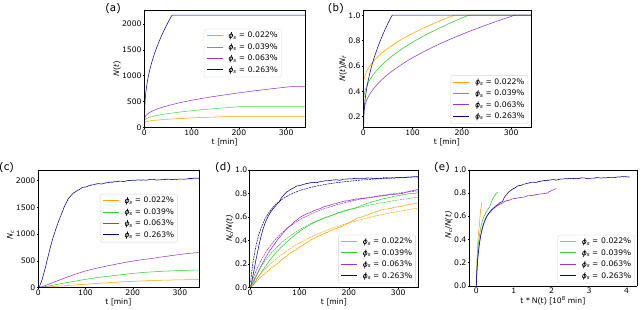}
\caption{\textbf{Time-dependent concentration in simulations.} Following experimental measurements we implement an increasing concentration in the simulations. The total number of particles in the system as a function of time for different final concentrations is shown in (a) and as a ratio of the final number of particles in (b). In (c) we show the total number of particles that are in a cluster (note the absence of significant jumps) while in (d) we show the relative number of particles. The solid lines are the data from simulations, the dashed lines best fits. In (e) we present the master curve.}
\label{fig:SI_Plots_TDC}
\end{figure*}

We find that a square-root fit $N(t) = N (t = 0) + b \sqrt{t}$, with two free parameters, approximates the experimental measurements well. We thus use these fits to obtain an expression for the concentration as function of time. This expression is then implemented in the simulation to increase the number of particles in the system over time as follows. While the concentration is less than the final concentration particles are added at the rate determined by the fit function. Once the final concentration is reached no more particles are added. This results in the $N(t)$ curves presented in Fig. \ref{fig:SI_Plots_TDC}a,b. Using these time-dependent concentrations we now turn towards studying the cluster dynamics. The results are presented in Fig. \ref{fig:SI_Plots_TDC}c-e. We now find a good quantitative agreement between experiments and simulations. This suggests that the time-dependent concentration was the major factor causing the large discrepancies between experiment and simulations mentioned in the main text. Note that the master curve we obtain (Fig. \ref{fig:SI_Plots_TDC}e) from the simulations is rather poor. The reason for this is found to be the naive implementation of the concentration increase. Namely, due to the hard cut-off once the final concentration, $N(t = t_{\rm fin}) = N_{\rm fin}$, is reached we find a small jump in the $N_c/N(t)$ curve exactly at the time $t_{\rm fin}$. This can be seen in Fig. \ref{fig:SI_Plots_TDC}d. When computing the master curve, the size of this jump is exaggerated, see Fig. \ref{fig:SI_Plots_TDC}e. Thus, for times before the earliest cut-off the curves for different concentrations fall onto a single master curve.  Note that this jump is not visible in the $N_c$ curve in \ref{fig:SI_Plots_TDC}c and is purely due to the hard cut-off which enters in Figs. \ref{fig:SI_Plots_TDC}d,e when the $y$-axis is normalized by dividing by $N(t)$. Furthermore, note that this is a temporary effect. We find that for very long times the curves converge towards a single master curve once again.

\newpage
\section*{Results for 90\textdegree{}-particles from simulations}
\begin{figure*}[h!]
\includegraphics[width=\textwidth]{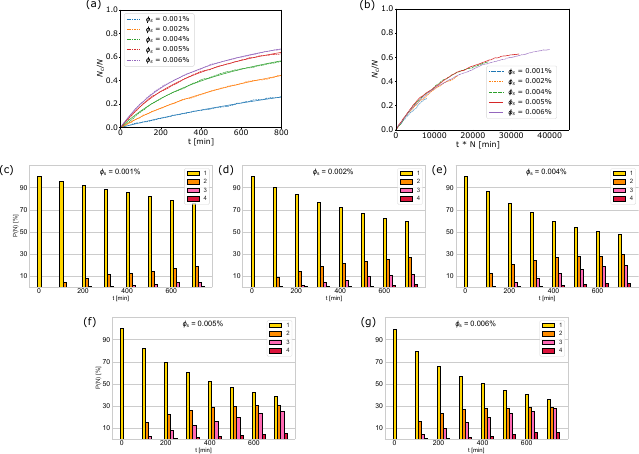}
\caption{\textbf{Results for 90\textdegree{}-particles from simulations.} In (a) the $N_c/N$ curves for different concentrations for particles with an opening angle of 90\textdegree{}. The corresponding master curve, found by rescaling time $t \to t * N$ is shown in (b).  In panels (c)-(g) we shown the histograms for the size of clusters over time for the different concentrations.}
\label{fig:SI_Plots_90}
\end{figure*}

We present here the results from the simulations of the active particles with an opening angle of 90\textdegree{} and the same arc-length as particles with an opening angle of 180\textdegree{} we simulated for Fig. 3 of the main text and Fig. \ref{fig:SI_Plots_TDC}. The results are thus different from the ones presented for simulations with an opening angle of 90\textdegree{} in the main text, where instead the \textit{cross section} was kept the same as the 180\textdegree{}-particles. Here, the cross section of the 90\textdegree{}-particles is thus bigger than the one of the 180\textdegree{}-particles with the same arc-length. We find the clustering curves $N_c/N$ presented in Fig. \ref{fig:SI_Plots_90}a. They are similar to the ones found for 180\textdegree{}-particles shown in Fig. 3d of the main text. The bigger cross-section of the 90\textdegree{} particles ``compensates'' for the smaller opening angle which, at the same cross-section, would result in clusters being more unstable. The corresponding master curve is shown in Fig. \ref{fig:SI_Plots_90}b. Finally, the cluster-size histograms for different concentrations are presented in Fig. \ref{fig:SI_Plots_90}c-g. The same quantities for 180\textdegree{}-particles of same arc-length are shown in Fig. \ref{fig:SI_Plots_180}.
\newpage
\section*{Simulations for different opening angles with same arc-length}
\begin{figure*}[h!]
\includegraphics[width=\textwidth]{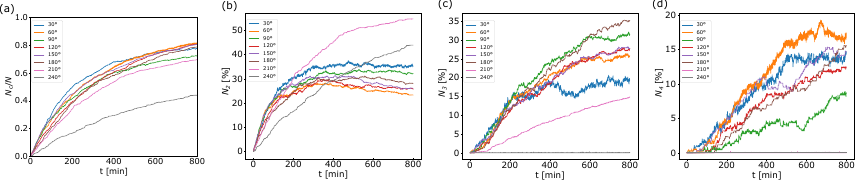}
\caption{\textbf{Simulation of particles with different opening angles but fixed arc-length}. We show in (a) the result for the $N_c/N$ curves. Apart from the largest angle (240\textdegree{}) the curves all angles considered are very similar. The percentage of two-, three-, and four-clusters over time is shown in panels (b)-(d), respectively. Large cluster sizes are rarely found.}
\label{fig:SI_Plots_DiffAngle_ArcLength}
\end{figure*}

Here we present the results for simulations of particles with different opening angles when the arc-length of the particles is fixed. That is, the smaller the opening angle, the larger the cross-section of the particles. This is different of the results presented in the main text (Fig. 5) where the particles considered have instead the same cross-section. We find that over a large range of opening angles the clustering dynamics is quite similar, see Fig. \ref{fig:SI_Plots_DiffAngle_ArcLength}. This is in stark contrast to the results for fixed cross-section where the clustering dynamics was significantly different for different opening angles.

\section*{Rescaling of the measured values for $\left(\tau_1^{180}/\tau_1^{90}\right)$}
We expect $\tau_1$ to scale inversely proportional with the speed, as faster particles encounter each other faster and hence assemble more quickly. Secondly, faster sedimentation to the glass surface increases the number of single particles and hence increases $\tau_1$. We also assume that this effect dominates over the faster increase in the particle density, as we use the density corrected value for $\tau_N$, which is $\tau_1$. These effects can be captured by rescaling the measured values for $\left(\tau_1^{180}/\tau_1^{90}\right)$ with the measured propulsion velocities for the two shapes (0.78 and 1.02 respectively) as well as with the inverse calculated sedimentation velocities (\cite{Tchen1954}). 
\\
\\
Once a sedimenting crescents has reached its terminal sedimentation velocity the following equilibrium holds: $F_{drag} = F_{g}$. From the work of Tchen et al.~\cite{Tchen1954} on the resistance experienced by particles with a similar shape, i.e. curved and elongated particles, we know that the fluid resistance, which here corresponds to the drag force $F_{drag}$, can be written as
\begin{equation}
\text{resistance} = 2R\chi_{0}*U*\zeta
\end{equation}
with $R$ the radius of the circle (see Fig. 1 of main text ), the angle $\chi_{0}$ which correspond to half the opening angle $\alpha$, $\zeta$ the frictional coefficient and $U$ the velocity of the flowing fluid which for a sedimenting particle is equal to the sedimentation velocity $v_{sed}$. Using these expressions we find for the sedimentation velocity 
\begin{equation}
v_{sed} = \frac{m'g}{\zeta*2R\chi_{0}}
\end{equation}
where $m'$ is the effective mass of the particle ($m'= m-V\rho$) and $\zeta$ can be written as \cite{Tchen1954}   
\begin{equation}
\zeta = \frac{m\pi\eta}{\ln(l/b_{0})+e}
\end{equation}
with $\eta$ the coefficient of viscosity, $l$ the half-arclength of the bent rod and $b_{0}$ the cross-sectional radius at the center of the particle which corresponds to half the thickness at that point. The kinematic shape factor $m$ and the dynamic shape factor $e$, which both depend on the angle $\chi_{0}$, can be taken from \cite{Tchen1954}.
\\
\\
For our 180\textdegree{} crescent with a thickness of $2 {\mu}{m}$, a radius of $4 {\mu}{m}$ and an arcleng of $4\pi$, using $m_{180}=3$ and $e_{180}=0.1$, we find $v_{sed}^{180}=1.14 {\mu}{m}/s$. While for our  90\textdegree{} crescent with a thickness of $1.5 {\mu}{m}$, a radius of $R=\sin({\pi/4})*8.5 {\mu}{m}$ and an arclength of $R*\pi/2$, and using $m_{90}=3.6$ and $e_{90}=0.8$ taken from Ref.~\cite{Tchen1954}, we obtain $v_{sed}^{90}=0.73 {\mu}{m}/s$. We can now use these velocities as well as the measured propulsion velocities for the two shapes ($0.78\pm0.08$~\si{\micro\meter\per\sec} and $1.02\pm0.03$~\si{\micro\meter\per\sec} respectively) to rescale the measured values for $\left(\tau_1^{180}/\tau_1^{90}\right)$ as follows:
\begin{equation}
\left(\frac{\tau_1^{180}}{\tau_1^{90}}\right)_{c}=\frac{\tau_1^{180}v_{prop}^{180}v_{sed}^{90}}{\tau_1^{90}v_{prop}^{90}v_{sed}^{180}}= \frac{2.78\cdot 0.78\cdot 0.73}{3.63\cdot 1.02\cdot 1.14}=0.38
\end{equation}
This is in perfect agreement with the value found in simulations and implies that hydrodynamic attractions seem negligible. 
\section*{Description of the Videos}
Video1\textunderscore Bent rod with $\alpha =\ang{180}$ in 1\% H2O2: \\
Bent rod with an opening angle of 180\textdegree{} self-propelling in a 1\% aqueous H2O2 solution. The particle moves concave-side leading.
\\
\\
Video2\textunderscore Bent rod with $\alpha=\ang{180}$ in 5\% H2O2:\\
Bent rod with an opening angle of 180\textdegree{} self-propelling in a 5\% aqueous H2O2 solution. The particle moves convex-side leading. 
\\
\\
Video3\textunderscore Bent rod with $\alpha=\ang{90}$ in 1\% H2O2:\\
Bent rod with an opening angle of 90\textdegree{} self-propelling in a 1\% aqueous H2O2 solution. The particle moves concave-side leading.
\\
\\
Video4\textunderscore pair formation\textunderscore $\alpha=\ang{180}$\textunderscore 1\% H2O2:\\
Formation of a 180\textdegree{} crescent-pair in a suspension of concave-side leading particles (1\% H2O2). Cluster exhibits rotational motion and almost no translational motion.
\\
\\
Video5\textunderscore rotating 3-particle cluster and Video6\textunderscore rotating 5-particle cluster:\\
Larger clusters of concave-side leading 180\textdegree{} crescents (1\% H2O2) also exhibit rotational motion and almost no translational motion.
\\
\\
Video7\textunderscore instable pair\textunderscore convex-side leading 180\textdegree{} crescents in 5\% H2O2:\\
The few pairs observed over the course of the experiment for convex-side leading 180\textdegree{} crescents (5\% H2O2) usually were not stable beyond max. 2 min.
\\
\\
Video8\textunderscore Cluster formation in simulations in small system:\\
We show results from a simulation of a small systems and for high density to illustrate the cluster formation process for 180\textdegree{} crescents. See the main text and the SM for details on the simulations. According to the mapping described in the SM, one second of the video corresponds to $\approx 4.8{\rm s}$ of physical time.
\\
\\
Video9\textunderscore Cluster formation in simulations in large system:\\
The system size presented here is the one used to obtain the results presented in this paper and corresponds to the system size of the experimental paper, according to the mapping explained in the SM. The surface area fraction here is $\phi_s = 0.05$\%. A two-cluster forms after about $15{\rm s}$ in the video, approximately at coordinates $(200, 600)$. One second of the video corresponds to $\approx 9.6{\rm s}$ of physical time.

\end{document}